\documentclass[10pt,twocolumn,letterpaper]{article}
\usepackage[accsupp]{axessibility} 
\usepackage[pagenumbers]{iccv}

\usepackage{pifont}
\newcommand{\cmark}{\color{black}{\ding{51}}}  
\newcommand{\xmark}{\color{gray}{\ding{55}}}  

\newcommand{\mypara}[1]{\vspace{0.1cm}\noindent{\bf{#1}}}
\usepackage{svg}
\usepackage{xcolor}
\usepackage{booktabs}
\usepackage{placeins}

\usepackage{siunitx}
\usepackage{amssymb} 
\usepackage[utf8]{inputenc} 
\usepackage{colortbl}  
\usepackage{makecell}  
\usepackage{pmboxdraw}      
\DeclareUnicodeCharacter{2713}{\check} 
\DeclareUnicodeCharacter{2717}{\cross} 
\usepackage{multirow}
\usepackage{soul} 
\usepackage{siunitx}
\usepackage{tikz}
\usetikzlibrary{positioning}
\usepackage{multicol} 
\usepackage{listings}
\usepackage{float}
\usepackage{wrapfig}
\definecolor{iccvblue}{rgb}{0.21,0.49,0.74}
\definecolor{sbblue}{HTML}{5273AD}
\definecolor{sbgrey}{HTML}{D3D3D3}
\definecolor{sborange}{HTML}{D78357}
\definecolor{sbgreen}{HTML}{5fA86C}
\definecolor{sbred}{HTML}{BD4C53}

\usepackage[pagebackref,breaklinks,colorlinks,allcolors=sbblue]{hyperref}

\newcommand{\vggsound}{\mbox{VGG}Sound}
\newcommand{\vggsounder}{\mbox{VGG}Sound\textit{er}}
\newcommand{\audioset}{AudioSet}
\newcommand{\flickrsoundnet}{Flickr-SoundNet}
\newcommand{\kineticssound}{Kinetics-Sound}

\usepackage[breakable]{tcolorbox}
\newcommand{\callout}[1]{%
    \begin{center}
    \begin{tcolorbox}[colframe=sbblue, colback=sbblue!30, coltext=sbblue!50!black, width=\columnwidth, left=.7mm, right=.7mm, top=1mm, bottom=1mm, boxrule=.5mm]
        #1
    \end{tcolorbox}
    \end{center}
}

\newcommand{\calloutt}[1]{%
    \begin{center}
    \begin{tcolorbox}[colframe=sbgrey, colback=sbgrey!30, coltext=sbgrey!20!black, width=\columnwidth, left=.7mm, right=.7mm, top=1mm, bottom=1mm, boxrule=.5mm]
        #1
    \end{tcolorbox}
    \end{center}
}

\newcounter{takeaway}
\newcommand{\takeaway}[1]{%
    \stepcounter{takeaway}
    \callout{\textbf{Takeaway\;\thetakeaway}\hspace{.5em}#1}
}

\newif\ifshowchanges
\showchangestrue 

\ifshowchanges
  
\else

\fi

\newcommand\mystrut{\rule[-1pt]{0pt}{.8em}}
\newtcbox{\hlbox}{%
    on line,
    colback=gray!20,
    colframe=white,
    boxrule=0pt,
    arc=1pt,
    left=-1pt, right=-1pt, top=-1.5pt, bottom=-1.5pt,
    fontupper={\ttfamily\small\mystrut}
}
\newcommand{\classname}[1]{\hlbox{#1}}
\definecolor{orangebg}{HTML}{F1CEB8}
\definecolor{bluebg}{HTML}{B7C7Df}
\definecolor{greenbg}{HTML}{BBDCC3}
\definecolor{greybg}{HTML}{999999}
\newtcbox{\orangebox}{%
    on line,
    colback=orangebg,
    colframe=white,
    boxrule=0pt,
    arc=1pt,
    left=-1pt, right=-1pt, top=-1.5pt, bottom=-1.5pt,
    fontupper={\ttfamily\small\mystrut}
}
\newtcbox{\bluebox}{%
    on line,
    colback=bluebg,
    colframe=white,
    boxrule=0pt,
    arc=1pt,
    left=-1pt, right=-1pt, top=-1.5pt, bottom=-1.5pt,
    fontupper={\ttfamily\small\mystrut}
}
\newtcbox{\greenbox}{
    on line,
    colback=greenbg,
    colframe=white,
    boxrule=0pt,
    arc=1pt,
    left=-1pt, right=-1pt, top=-1.5pt, bottom=-1.5pt,
    fontupper={\ttfamily\small\mystrut}
}
\newtcbox{\greybox}{
    on line,
    colback=greybg,
    colframe=white,
    boxrule=0pt,
    coltext=white,
    arc=1pt,
    left=-1pt, right=-1pt, top=-1.5pt, bottom=-1.5pt,
    fontupper={\ttfamily\small\mystrut}
}

\usepackage{scalerel,graphicx,xparse}
\NewDocumentCommand\boar{}{\raisebox{-4px}{\includegraphics[height=20px]{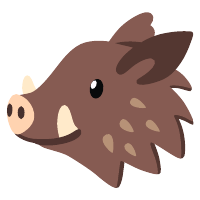}}}

\def\paperID{3641} 
\def\confName{ICCV}
\def\confYear{2025}

\title{\boar{} \vggsounder{}: Audio-Visual Evaluations for Foundation Models}

\author{
    Daniil Zverev\textsuperscript{1,}\thanks{equal contribution} \hspace{6pt}
    Thaddäus Wiedemer\textsuperscript{2,3,4,*}\hspace{6pt}
    Ameya Prabhu\textsuperscript{2,3} \\[0.3em]
    Matthias Bethge\textsuperscript{2,3} \hspace{6pt}
    Wieland Brendel\textsuperscript{3,4} \hspace{6pt}
    A. Sophia Koepke\textsuperscript{1,2,3} \\[0.7em]
    \textsuperscript{1}Technical University of Munich, MCML \hspace{10pt}
    \textsuperscript{2}University of Tübingen  \hspace{10pt} \textsuperscript{3}Tübingen AI Center \\[0.3em]
    \textsuperscript{4}MPI for Intelligent Systems, ELLIS Institute Tübingen \\[0.4em]
    {\small \tt \{daniil.zverev,a-sophia.koepke\}@tum.de},
    {\small \tt \{thaddaeus.wiedemer,ameya.prabhu\}@uni-tuebingen.de}
}

\begin{document}
\maketitle

\begin{abstract}
The emergence of audio-visual foundation models underscores the importance of reliably assessing their multi-modal understanding. The \vggsound{} dataset is commonly used as a benchmark for evaluating audio-visual classification. However, our analysis identifies several limitations of \vggsound{}, including incomplete labelling, partially overlapping classes, and misaligned modalities. These lead to distorted evaluations of auditory and visual capabilities. To address these limitations, we introduce \vggsounder{}, a comprehensively re-annotated, multi-label test set that extends \vggsound{} and is specifically designed to evaluate audio-visual foundation models. \vggsounder{} features detailed modality annotations, enabling precise analyses of modality-specific performance. Furthermore, we reveal model limitations by analysing performance degradation when adding another input modality with our new modality confusion metric.
Our dataset and project page are available at \url{https://vggsounder.github.io/}.
\end{abstract}

\section{Introduction}
\begin{figure}[t]
    \centering
    \includegraphics[width=\linewidth]{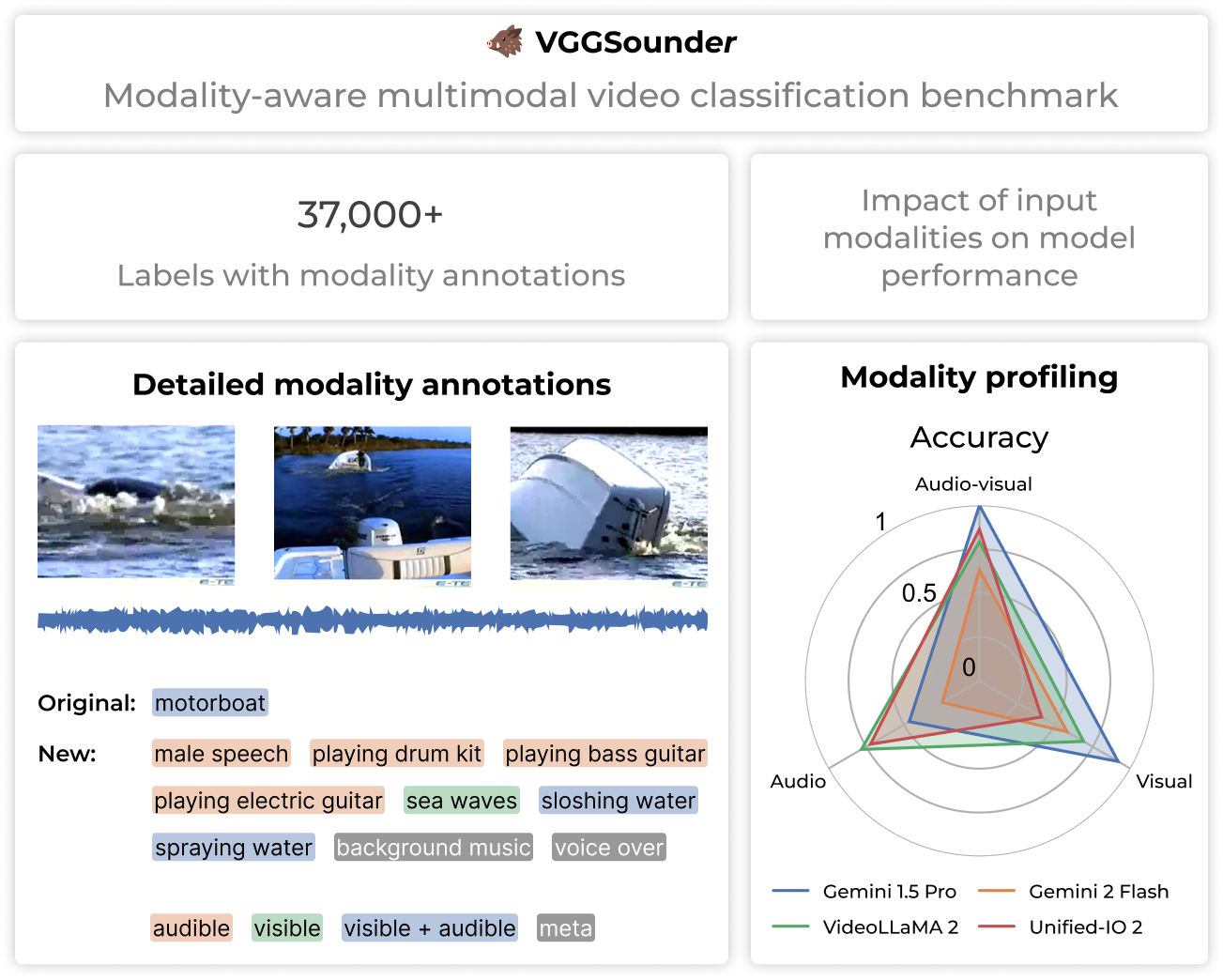}
    \vspace{-1em}
    \caption{
        \textbf{We introduce \vggsounder{}, a multi-label audio-visual classification benchmark with modality annotations}.
        We extend the original \vggsound{} test set with human-annotated \orangebox{audible}, \greenbox{visible}, and \bluebox{visible$+$audible} labels. We add \greybox{meta} labels for common confounders, such as background music. We benchmark eleven recent audio-visual models on \vggsounder{}.
        It enables selective analysis of a model’s auditory and visual capabilities on classes relevant for the queried modality. 
    }
    \label{fig:overview}
   \vspace{-0.7em}
\end{figure}

Rigorous evaluation benchmarks have been instrumental in assessing the effectiveness of audio-visual models~\citep{gong2022contrastive, mo2024unveiling, lin2024siamese, kim2024equiav}. Specifically, multi-modal foundation models integrating visual and auditory data aim to achieve a holistic understanding of audio-visual content. However, the field lacks large-scale modality-aware classification benchmarks with ground-truth annotations indicating whether each label is visible, audible, or both. Such annotations would allow detailed evaluations of multi-modal model capabilities. To address this gap, we introduce \vggsounder{}, an enhanced version of the widely-used audio-visual classification dataset \vggsound{}~\cite{chen2020vggsound}, which facilitates modality-aware evaluation of audio-visual foundation models.

\vggsound{} has several notable limitations. First, its data is inherently multi-label; for instance, a single sample may simultaneously include labels such as \classname{playing drum kit} and \classname{playing acoustic guitar} when multiple instruments are present. Additionally, evaluating how different modalities contribute to model performance becomes difficult without explicit modality annotations, as some labels are either not visually present or not audible (e.g., certain instruments might only be audible but not visible in advertisements). Moreover, overlapping label classes present another challenge; for example, the \classname{orchestra} label often coincides with labels for individual instruments. These issues result in systematic under-evaluation of multi-modal audio-visual foundation models.

To overcome these limitations, we present \vggsounder{}, an improved benchmark inspired by similar advancements in other domains~\citep{beyer2020we, gema2024mmlu}. We re-annotate the dataset to create a comprehensive multi-label classification setting by collecting detailed annotations for each sample, including (1)~additional classes present, (2)~explicit modality annotations to label modality misalignment, (3)~metadata indicating the presence of background music, voice-over, or static images, and (4)~merging of classes to address overlapping classes. Consequently, \vggsounder{} provides a robust, foundation-model-ready benchmark enabling structured analysis of whether models rely on audio or visual cues. Furthermore, we include meta-labels (e.g., background music, voice-over, or static images) to easily filter out unreliable labels during evaluation. Utilising \vggsounder{}, we evaluate audio-visual foundation models, demonstrating their poor performance on our benchmark. We find that the state-of-the-art, closed-source Gemini models consistently rely exclusively on the visual modality. We measure that effect with the modality confusion, i.e.\ when models get distracted by an additional input modality, which exposes the unsuccessful merging of modalities. These findings highlight the importance of the audio-focused \vggsounder{} benchmark as a critical tool for accurately assessing audio-visual foundation models.

\callout{%
We make the following contributions:
\begin{enumerate}
    \item We illustrate limitations of \vggsound{} in \cref{sec:vggsound-issues}.
    \item We curate \vggsounder{} with multi-modal human annotations for multi-label classification in \cref{sec:vggsounder}.
    \item We evaluate state-of-the-art audio-visual models, observing differences between embedding models and autoregressive foundation models in \cref{sec:experiments}.
    \item We propose new metrics to quantify the negative impact of using multiple input modalities in \cref{sec:experiments}.
\end{enumerate}
}

\section{Related work}
\label{sec:related-work}

\mypara{Audio-visual learning}
Many prior works consider audio-visual tasks that include sound source localisation and separation~\citep{owens2018audio,arandjelovic2018objects,gao2019co,chen2021localizing,Afouras20b,afouras2021selfsupervised,qian2020multiple,xu2020cross,tzinis2020into,zhu2022v,zhao2018sound,zhao2019sound,mo2022localizing}, event localisation~\citep{tian2018audio,lin2019dual,wu2019dual,lin2020audio-visual}, audio-visual question answering~\cite{yang2022avqa,li2022learning,yun2021pano,ma2024look}, audio-visual synchronisation~\citep{chen2021audio,chung2016out,ebeneze2021detection,khosravan2019attention,iashin2024synchformer,Iashin_2022_BMVC}, audio synthesis using visual information~\citep{zhou2019vision, goldstein2018guitar,su2020multi,gan2020foley,narasimhan2021strumming,koepke2020sight,su2021does,su2024vision,chen2024video,koepke2019visual}, or audio-driven face image synthesis~\citep{wiles2018x2face,jamaludin2019you,aneja2023facetalk}.
Audio-visual data has also been leveraged for speech-related tasks, including speech and speaker recognition~\citep{afouras2020asr,afouras2018deep,nagrani2020disentangled}, or the spotting of spoken keywords~\citep{momeni2020seeing,Prajwal2021}.

\noindent Furthermore, the natural alignment between audio and video has been exploited to learn improved audio-visual embeddings for downstream tasks~\citep{owens2016ambient,owens2018learning,alwassel2019self,patrick2020multi,korbar2018cooperative,aytar2016soundnet,chen2021distilling,asano2020labelling,nagrani2021attention,xiao2020audio-visual,cheng2020look,cheng2024videollama}.
Using both modalities jointly generally leads to performance boosts over using one modality in isolation.
We examine this observation closely and aim to evaluate the effective use of multiple input modalities for the video classification task. To enable this, we propose — to the best of our knowledge, the first multi-label video classification benchmark that includes per-modality annotations for every sample (see \cref{tab:bench-comparison}).

\begin{table}
    \centering
    \resizebox{\linewidth}{!}{
        \begin{tabular}{lrrccl}
        \toprule
        \textbf{Dataset} & \textbf{\# Clips} & \textbf{\# Classes} & 
        \makecell[b]{\textbf{Multi-}\\\textbf{label}} & 
        \makecell[b]{\textbf{Modality}\\\textbf{labels}} & 
        \makecell[lb]{\textbf{Annotation}\\\textbf{pipeline}} \\
        \midrule
        \flickrsoundnet{}~\cite{aytar2016soundnet} & 2M & - & \xmark & \xmark & - \\
        \kineticssound{} ~\cite{arandjelovi2017look} & 18.8k & 34 & \xmark & \xmark & MTurk \\
        \audioset{}~\cite{gemmeke2017audio} & 2.1M& 537 & \cmark & \xmark & Manual \\
        ├─ AVE~\cite{tian2018audio} & 4k & 28 & \cmark & \xmark & Manual \\
        ├─ VEGAS~\cite{Zhou_2018_CVPR} & 132k & 10 & \xmark & \xmark & MTurk \\
        └──┴─ Visually Aligned Sounds~\cite{9151258} & 13k & 8 & \xmark & \xmark & MTurk \\
        Perception Test ~\cite{pătrăucean2023perceptiontestdiagnosticbenchmark}& 11.6k & 16 & \cmark & \cmark & Manual \\ 
        Epic-Sounds ~\cite{huh2025epicsoundslargescaledatasetactions} &  0.7k &  44 &  \cmark & \cmark & Manual \\
        \vggsound{}~\cite{chen2020vggsound} & 200k & 309 & \xmark & \xmark & Classifiers+Manual \\
        ├─ \vggsound{}-Sparse~\cite{Iashin_2022_BMVC} & 7.1k & 12 & \xmark & \xmark & Manual \\
        ├─ Visual Sound~\cite{viertola2024temporallyalignedaudiovideo} & 91k & 309 & \xmark & \xmark & ImageBind~\citep{girdhar2023imagebind} \\
        └─ \vggsounder{}\textsuperscript{*} & 15k  & 309 & \cmark & \cmark & MTurk \\
        \bottomrule
        \multicolumn{6}{r}{\footnotesize\textsuperscript{*}VGGSounder v0.1.6 has 15.4k samples.}
        \end{tabular}
    }
    \caption{Comparison of audio-visual classification benchmarks.}
    \label{tab:bench-comparison}
   \vspace{-0.7em}
\end{table}

\mypara{Audio-visual foundation models} Recently, multi-modal general-purpose models have emerged that can handle diverse downstream tasks without task-specific finetuning -- also referred to as multi-modal foundation models. For instance, images or language were used as the bridge between modalities including audio, image, and text~\cite{girdhar2023imagebind,zhu2023languagebind}.
Building on this, PandaGPT~\cite{suPandaGPTOneModel2023} leverages Vicuna~\cite{chiang2023vicuna} and ImageBind's embedding space to train a general multi-modal model exclusively on image-text pairs. 
Unified-IO 2~\cite{luUnifiedIO2Scaling2023c} employs universal tokenisation to process audio, video, and text. VideoLLaMA2~\cite{cheng2024videollama} uses a Spatial-Temporal Convolution connector in the visual branch before projecting audio and visual information into the LLM input space. The recently introduced Ola model~\cite{liuOlaPushingFrontiers2025} advances omni-modal processing through progressive modality alignment, using video to bridge audio and visual information. The Gemini models~\cite{team2023gemini} are closed-source multi-modal models that achieve impressive performance on diverse downstream tasks. We use \vggsounder{} to benchmark the audio-visual capabilities of the aforementioned models.

\begin{figure*}[t]
    \centering
    \includegraphics[width=0.95\linewidth]{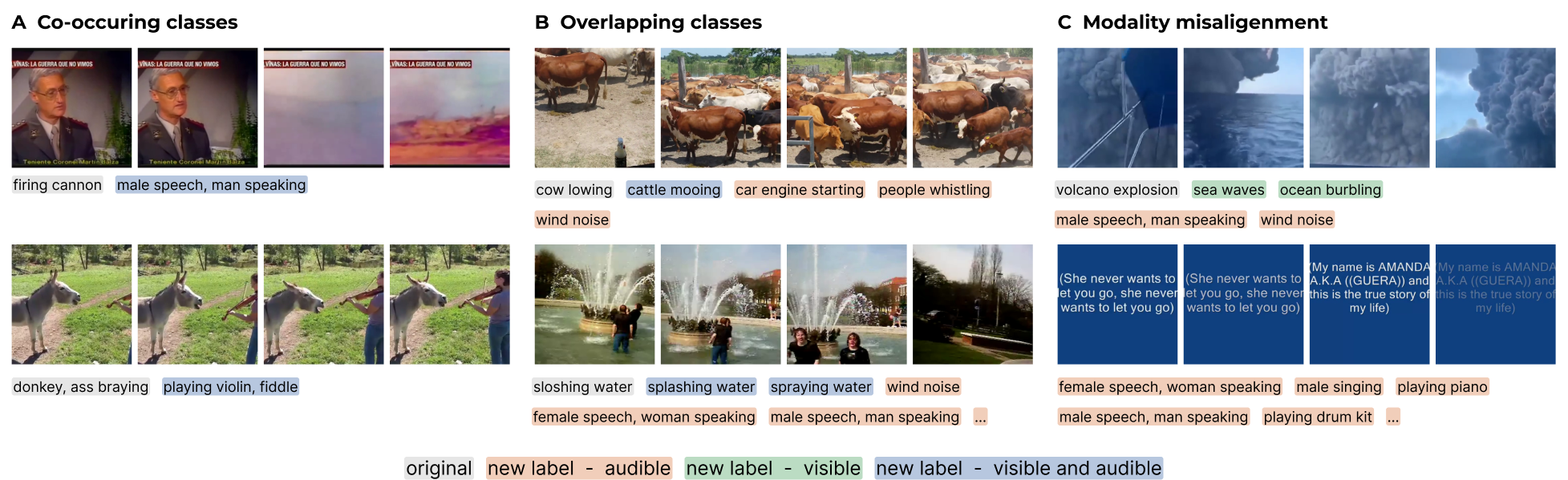}
    \caption{
        \textbf{Limitations of VGGSound.} We show video frames from videos in the VGGSound test set along with their annotated label (grey) to demonstrate various limitations. 
        \textbf{A}. VGGSound samples are labelled with a single class, yet many videos contain multiple distinct classes.
        \textbf{B}. Additionally, many classes partially overlap or are ambiguous. 
        \textbf{C}. Some samples are labelled with classes that are not present in one of the modalities (i.e., the labelled class is not visible or audible).
    }
    \label{fig:vggsound-issues}
   \vspace{-0.5em}
\end{figure*}

\mypara{Audio-visual classification benchmarks} Audio-visual classification is distinct from general video classification (e.g.\ on YouTube-8M~\citep{abu2016youtube}), as classes typically cover both audible and visible actions or events.
Commonly used datasets for audio-visual classification include \kineticssound{}~\citep{arandjelovi2017look} sourced from the Kinetics dataset~\citep{kay2017kinetics}, \flickrsoundnet{}~\citep{aytar2016soundnet} scraped from Flickr, and \audioset{}~\citep{gemmeke2017audio} and \vggsound{}~\citep{chen2020vggsound}, both sourced from YouTube.

\noindent \kineticssound{} features manual labels of human actions, but covers only \num{34} classes.
\flickrsoundnet{} is much larger, but only a small subset is labelled.
Similarly, only a small fraction of the roughly \si{2}{M} \audioset{} samples are annotated and have aligned audio and video.

\noindent In contrast, \vggsound{} ensures audio-visual correspondence for around \num{200000} samples and was curated with an automatic pipeline involving class-list generation, and auditory and visual content verification.
The visual verification step ensures that a class is represented in the centre frame.

\noindent The VEGAS dataset ~\citep{Zhou_2018_CVPR} provides better quality assurances for a small subset of \audioset{} with only \num{10} classes.
Visually Aligned Sounds~\citep{9151258} subsamples VEGAS and \audioset{} after human verification, and Visual Sound~\citep{viertola2024temporallyalignedaudiovideo} subsamples \vggsound{} using a multi-modal embedding model, both aiming for high audio-visual correspondence.
Similarly, \vggsound{}-Sparse~\citep{Iashin_2022_BMVC} is a subset of \vggsound{} with a focus on temporally and spatially sparse synchronisation signals (e.g., short loud noises). Perception test ~\cite{pătrăucean2023perceptiontestdiagnosticbenchmark} offers multi-label rich modality annotations and is specifically designed for LLM benchmarking, however it has a very small number of classes. Epic-Sounds ~\cite{huh2025epicsoundslargescaledatasetactions} also comes with modality annotation, but offers only ego-centric related classes ~\cite{huh2025epicsoundslargescaledatasetactions}.

\noindent Overall, \vggsound{} strikes the best balance between size, generality, and annotations, making it a common benchmark for audio-visual classification.
We update \vggsound{} to sustain its usability for the development of the next generation of multi-modal foundation models.

\section{Limitations of \vggsound{}}\label{sec:vggsound-issues}
\noindent Since we are interested in the \vggsound{} dataset for benchmarking audio-visual multi-modal models, our analysis focuses on the \vggsound{} test set,\footnote{Although these issues most likely also apply to the training set.}
which consists of \num{15446} video clips, each \num{10}{s} long and labelled with one of \num{309} classes.
We qualitatively identify several limitations of the \vggsound{} annotations outlined below and in \cref{fig:vggsound-issues}.

\mypara{Co-occurring classes}
While \vggsound{}'s visual verification aimed to minimise multiple classes co-occurring in a clip, we find that most samples nevertheless clearly contain multiple classes, see \cref{fig:vggsound-issues}A.
In some cases, classes are temporally separated, e.g., showing \classname{male speech, man speaking} and then cutting to footage of \classname{firing cannon}.
Most often, classes co-occur at the same time, sometimes for the entire duration of the video clip.
Overlapping classes are often related, such as different instruments in a band or orchestra, but can also be entirely unrelated, e.g., \classname{donkey, ass braying} co-occurring with \classname{playing violin}. As additional empirical evidence, we provide a co-occurrence matrix computed using CAV-MAE~\citep{gong2022contrastive}, a state-of-the-art audio-visual model, in \cref{sec:additional-quantitative-analysis}.

\mypara{Overlapping classes}
The issue of co-occurring classes is exacerbated by many of the \num{309} automatically generated classes partially overlapping in their definition, as illustrated in \cref{fig:vggsound-issues}B.
We found two pairs of synonymous classes: \classname{timpani} and \classname{tympani} and \classname{dog barking} and \classname{dog bow-wow}.
Additionally, some classes are strict subclasses of others, such as the gender-specific versions of \classname{cattle mooing}: \classname{cow lowing} and \classname{bull bellowing}; or the more specific variants of \classname{people eating}: \classname{people eating noodle} and \classname{people eating apple}.
Finally, several classes commonly appear together, such as  \classname{playing snare drum} which is often played as part of a \classname{drum kit}, or semantically similar concepts: \classname{running electric fan} and \classname{air conditioning noise}, and \classname{sloshing water} and \classname{splashing water}.

\begin{figure*}[t]
    \centering
    \includegraphics[width=0.95\textwidth]{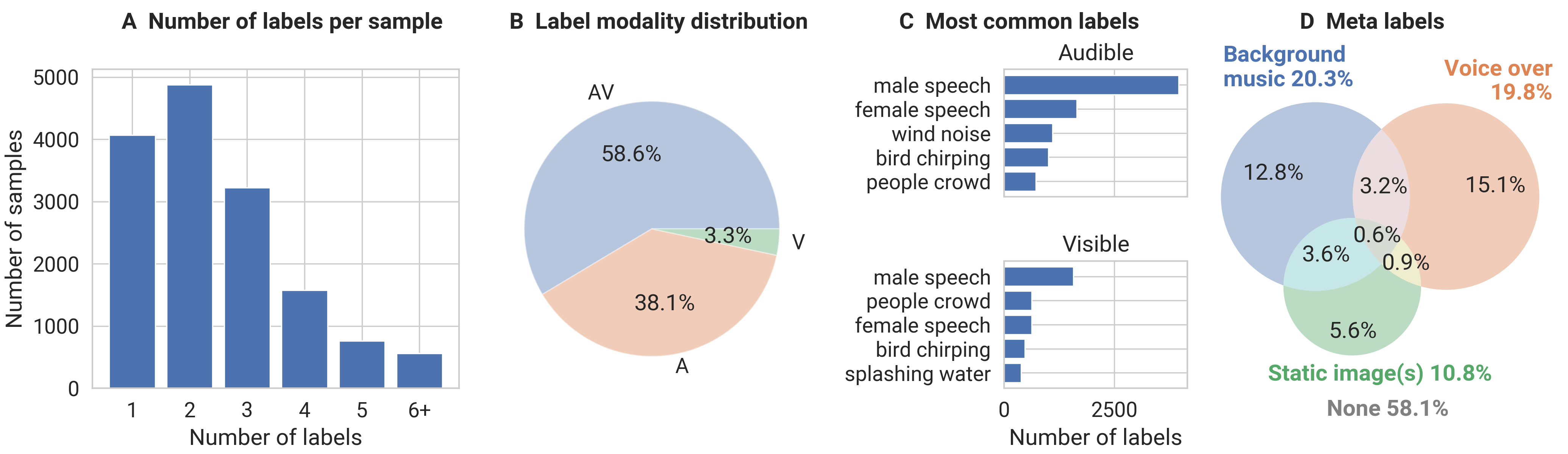}
    \caption{
        \textbf{Overview of \vggsounder{}}.
        \textbf{A}. Most samples contain more than one label.
        \textbf{B}. More than a quarter of labels are audible but not visible. In contrast, only a tiny fraction is visible but not audible.
        \textbf{C}. Speech and bird sounds are the most common classes; more details can be found in \cref{sec:class-label-frequency-in-vggsounder}.
        \textbf{D}. Forty percent of the samples contain some combination of \classname{background music}, \classname{voice over}, and \classname{static image(s)}, making the classification task significantly harder.
    }\label{fig:stats}
\vspace{-0.7em}
\end{figure*}

\mypara{Modality misalignment}
Despite \vggsound{}'s auditory and visual content verification, we find that many of the annotated classes are not visible or not audible, as shown in \cref{fig:vggsound-issues}C. 
A large fraction of videos contains background music, voice-over and narration, or other background sounds like \classname{bird chirping, tweeting} or \classname{cricket chirping} without a visible source.
Similarly, some videos contain visible but inaudible cues for classes like \classname{sea waves}.
Static images and slide shows accompanied by music or other sounds are other frequent sources of misaligned modalities.
Finally, some classes are misaligned by definition: \classname{wind noise} is only audible and not visible.

\noindent We find that, \SI{48.43}{\%} of the original \vggsound{} test samples have misaligned modalities. This finding challenges the widely held assumption that \vggsound{} has strong modality alignment~\citep{gong2022uavm, lee2021acav100mautomaticcurationlargescale}.

\noindent Other datasets, such as Visually Aligned Sounds, Visual Sound, and \vggsound{}-Sparse (see \cref{sec:related-work}) omitted samples with misaligned modalities.
In contrast, we contend that inaudible or invisible cues are common in natural videos and should be considered when benchmarking multi-modal models. We, therefore, place particular emphasis on crafting reliable modality annotations for all samples, allowing users to evaluate models on samples that guarantee modality alignment, and on those where classes are only visible or only audible (see \cref{tab:bench-comparison}).

\takeaway{\vggsound{} suffers from several issues: class co-occurrence not captured by single labels, {overlapping class definitions}, and {modality misalignment}, see \cref{fig:vggsound-issues}.}

\section{Building \vggsounder{}}\label{sec:vggsounder}
We propose a series of fixes for \vggsound{}'s issues, ultimately resulting in the updated \vggsounder{}  benchmark. We are not the first aiming to future-proof an existing benchmark:
\cite{beyer2020we} analysed shortcomings of ImageNet~\citep{deng2009imagenet}, ultimately proposing switching to a multi-label classification task with additional manual labels.
\cite{gema2024mmlu} similarly re-annotated samples in MMLU~\citep{hendryckstest2021,hendrycks2021ethics} to fix labelling errors.
Both works inspire our approach to improve \vggsound{}. 

\noindent To deal with co-occurring classes, we switch to a multi-label classification setting.
This effectively handles most overlapping class definitions: a strong model can assign a high probability to multiple classes, even if they partially overlap.
This also allows us to ensure that synonymous classes, as well as subclasses and their superclass, always appear together in the ground-truth labels.

\noindent To deal with modality misalignment, we add a modality annotation to each label.
For example, we can label a video as containing \classname{people clapping} in the audio and containing \classname{playing volleyball} in the audio and video data.
We also add meta-labels to indicate whether a sample contains \classname{background music}, \classname{voice over}, or \classname{static image(s)} to optionally treat these cases separately during evaluation.

\noindent We employ a pipeline similar to \cite{beyer2020we} to annotate multiple labels per sample, which we summarise below. Further details can be found in \cref{sec:vggsounder-relabelling-vggsound}.

\mypara{Collecting proposals}
We create a \emph{gold standard} reference set by labelling a small, randomly selected subset of \vggsound{} test samples with four in-house computer vision experts.
The interface used for this first labelling step is shown in \cref{sec:vggsounder-relabelling-vggsound}.
We extend the subset until each class is covered at least once, leading to a final size of \num{417} samples.
Labels from different annotators are merged via a simple majority vote.

\noindent Given the gold standard set, we want to find a solid strategy for automatically generating label proposals which are shown to the humans labelling the test set. This should have a recall greater than \SI{90}{\%} while maximising precision compared to the gold standard labels to produce a small set of proposals with good label coverage.
Our final strategy combines predictions from several state-of-the-art models with a manual heuristic to obtain \si{93}{\%} recall for an average of \num{30} proposals per sample.

\mypara{Human labelling}
We use Amazon Mechanical Turk to re-annotate the entire \vggsound{} test set.
For each sample, we ask annotators to indicate whether the video contains \classname{background music}, \classname{voice over}, or \classname{static image(s)}.
Then, annotators are asked to indicate for each label proposal whether the class is \classname{audible} and/or \classname{visible}.
Finally, annotators can add a class if it is missing from the proposals.
Annotators were paid the US minimum wage;
\noindent We let annotators label the samples in batches of 20, each containing two gold standard samples as catch trials.
We reject and re-annotate all batches with a catch trial $F_1$-score below \si{25}{\%}.
In addition to MTurk, we obtain many labels from in-house expert annotations. 

\mypara{Final labels}
We merge all obtained annotations using Dawid--Skene \cite{sinha2018fastdawidskenefastvote}.
Additionally, we add synonymous classes and superclasses for a given subclass, e.g., we add \classname{cattle mooing} whenever \classname{cow lowing} is in the set of labels. The labelling interface is shown in \cref{sec:vggsounder-relabelling-vggsound}.

\takeaway{We develop \vggsounder{}: A multi-label video classification benchmark extending \vggsound{} with human-annotated multi-labels, modality annotations, and meta-labels as summarised in \cref{fig:stats}.}

\section{Benchmarking audio-visual models}\label{sec:experiments}
We use \vggsounder{} to benchmark four popular audio-visual embedding models and seven foundation models, and analyse their auditory and visual capabilities.

\mypara{Models}
We evaluate the audio-visual \emph{embedding models} CAV-MAE~\citep{gong2022contrastive}, DeepAVFusion~\citep{mo2024unveiling}, AV-Siam~\citep{lin2024siamese}, and Equi-AV~\cite{kim2024equiav}. Those were finetuned on VGGSound. 

\noindent We benchmark several models from the closed-source Gemini family~\cite{team2023gemini} in a zero-shot evaluation protocol. Furthermore, we use LLM-assisted evaluation to evaluate the following four open-source autoregressive \emph{foundation models}: VideoLLaMA-2~\cite{cheng2024videollama}, Unified-IO-2~\cite{luUnifiedIO2Scaling2023c}, Panda-GPT~\cite{suPandaGPTOneModel2023}, and Ola~\cite{liuOlaPushingFrontiers2025}. 
All models are evaluated in three modes: using unimodal-audio, unimodal-visual, or multi-modal (audio and visual) inputs. Further details about models and their evaluation are provided in \cref{sec:supp:models}.

\mypara{Metrics}
To benchmark the models on \vggsounder{}, we use multi-label classification metrics.
For embedding models, all metrics are computed for the top-$k$ predictions, with $k \in \{1, 3, 5, 10\}$.
In contrast, prompting foundation models yields an unordered set of class predictions of varying size, and we compute only a single metric using the entire set.
As a result, metrics are not directly comparable between embedding and foundation models.
To get a sense of their relative performance, we report metrics for embedding models for $k=1$ in the main text, matching the median number of predictions per sample for the foundation models.

\noindent For open-source models such as VideoLLaMA-2, Ola, Unified-IO-2, and Panda-GPT, we employ LLM-assisted evaluation~\citep{maazVideoChatGPTDetailedVideo2024, xuPointLLMEmpoweringLarge2024}, in which the Qwen-3 model~\citep{yangQwen3TechnicalReport2025} is tasked to assess the correspondence between model outputs and target classes. Closed-source models from the Gemini family are evaluated by providing the full list of 309 classes as input. Further details on the evaluation procedures and exact prompts are provided in \cref{sec:input-prompts}.

\noindent \emph{Subset accuracy} compares the predicted label set to the ground-truth label set and reports the fraction of samples for which they match. This is our strictest metric.

\noindent $F_1$-\emph{score} is the harmonic mean of precision and recall. It is strictly larger than the subset accuracy.

\noindent \emph{Hit} reports the fraction of samples for which \emph{any} of the predicted labels are part of the ground-truth label set.
This is the most lenient metric which is strictly larger than the $F_1$-score.
We include this metric for ease of comparison to the ``ReaL-Accuracy'' used in \cite{beyer2020we}.

\noindent All metrics are computed, on a subset of \vggsounder{} without \classname{background music} labels, separately for each input modality (audio, video, and audio-visual) and label modality. We use lowercase symbols a, v, and av to indicate the input modality: audio-only, visual-only, or audio-visual inputs, respectively. For label modality, we use uppercase symbols \emph{A}, \emph{V}, and \emph{AV}, referring to the subsets of the benchmark with audible, visible, and audio-visual labels. We further include \emph{$A\neg V$} (audible but not visible) and \emph{$V\neg A$} (visible but not audible) to analyze unaligned cues. For clarity, we define shorthand notations such as $a = a(A)$ to denote the model’s performance on the audible subset \emph{A} using only audio input. Analogously, $v = v(V)$ and $av = av(AV)$ refer to video-only and audio-visual performance on their respective label subsets. Furthermore, we use micro-aggregation to balance the contribution from each class.  

\begin{table*}[t]
\centering
\resizebox{\textwidth}{!}{
\begin{tabular}{l S[table-format=2.2] S[table-format=2.2] S[table-format=2.2] S[table-format=2.2] S[table-format=2.2] S[table-format=2.2] S[table-format=2.2] S[table-format=2.2] S[table-format=2.2] S[table-format=2.2] S[table-format=2.2] S[table-format=2.2] S[table-format=2.2] S[table-format=2.2]}
\toprule
& \multicolumn{3}{c}{\textbf{Subset Accuracy} \small$\uparrow$} & \multicolumn{5}{c}{$\mathbf{F_1}$ \small$\uparrow$} & \multicolumn{3}{c}{\textbf{Hit} \small$\uparrow$} & \multicolumn{3}{c}{$\mathbf{\mu}$ \small$\downarrow$} \\
\cmidrule(lr){2-4} \cmidrule(lr){5-9} \cmidrule(lr){10-12} \cmidrule(lr){13-15}
\textbf{Model} & \emph{a} & \emph{v} & \emph{av} & \emph{a} & \emph{v} & \emph{av} & \emph{a($A\neg V$)} & \emph{v($V\neg A$)} & \emph{a} & \emph{v} & \emph{av} & \emph{$\mu_A$} & \emph{$\mu_V$} & \emph{$\mu_{A\cap V}$} \\

\midrule[0.75pt]
\multicolumn{15}{l}{\emph{Embedding Models}}\\
\addlinespace[2pt] 

CAV-MAE & \bfseries 17.94 & 24.21 & \bfseries 29.61 & \bfseries 39.53 & 38.87 & \bfseries 47.24 & \bfseries 21.69 & 23.56 & \bfseries 67.07 & 56.37 & \bfseries 67.51 & \bfseries 4.07 & 4.93 & 0.85 \\
DeepAVFusion & 13.89 & 13.83 & 26.16 & 29.06 & 23.29 & 41.81 & 15.67 & 13.65 & 49.40 & 33.80 & 59.79 & 4.21 & \bfseries 3.42 & \bfseries 0.26 \\
Equi-AV & 15.80 & 13.57 & 24.40 & 33.65 & 22.93 & 39.33 & 19.02 & 14.39 & 57.09 & 33.25 & 56.21 & 7.77 & 6.28 & 1.51 \\
AV-Siam & 17.40 & \bfseries 24.99 & 27.86 & 38.12 & \bfseries 39.77 & 43.73 & 20.41 & \bfseries 23.77 & 64.69 & \bfseries 57.68 & 62.49 & 10.76 & 8.55 & 3.98 \\

\midrule
\multicolumn{15}{l}{\emph{Closed-source Foundation Models}}\\
\addlinespace[2pt]

Gemini 1.5 Flash & 2.36 & 18.55 & 20.24 & 14.21 & 42.02 & 46.82 & 17.06 & 30.96 & 30.75 & 51.79 & 62.30 & 13.25 & 4.12 & 1.45 \\
Gemini 1.5 Pro & \bfseries 3.48 & \bfseries 27.18 & \bfseries 26.91 & \bfseries 19.38 & \bfseries 54.90 & \bfseries 57.20 & \bfseries 20.02 & \bfseries 32.14 & \bfseries 33.43 & \bfseries 73.27 & \bfseries 78.36 &  2.65 & \bfseries 4.03 & \bfseries 0.52 \\
Gemini 2.0 Flash & 2.43 & 16.31 & 15.48 & 12.86 & 38.67 & 40.50 & 7.73 & 26.45 & 18.86 & 47.01 & 50.56 & \bfseries 2.59 & 4.79 & 1.02 \\

\midrule
\multicolumn{15}{l}{\emph{Open-source Foundation Models}}\\
\addlinespace[2pt] 

VideoLLaMA 2 & \bfseries 16.25 & \bfseries 23.88 & 27.70 & 40.76 & \bfseries 50.28 & \bfseries 53.67 & 25.55 & \bfseries 38.01 & \bfseries 58.15 & \bfseries 53.22 & 59.92 & 14.51 & 5.43 & 3.11 \\
Unified-IO 2 & 15.27 & 14.65 & \bfseries 29.71 & 37.69 & 30.51 & 50.10 & 27.43 & 21.30 & 54.43 & 32.18 & \bfseries 63.36 & 10.80 & \bfseries 4.64 & \bfseries 1.84 \\
PandaGPT & 3.40 & 5.35 & 6.30 & 18.58 & 19.39 & 21.55 & 17.66 & 18.73 & 19.72 & 17.10 & 18.74 & \bfseries 9.45 & 6.33 & 3.03 \\
OLA & 15.99 & 10.54 & 19.00 & \bfseries 48.47 & 26.52 & 46.04 & \bfseries 45.53 & 17.35 & 57.50 & 24.84 & 50.20 & 17.34 & 6.23 & 2.85 \\

\bottomrule
\end{tabular}
}
\caption{\textbf{Audio-visual video classification results on \vggsounder{}}. We report multi-label classification metrics (subset accuracy, $F_1$-score, Hit accuracy, modality confusion $\mu$) on \classname{background music} free subset for audio- \emph{$a(A)$}, visual - \emph{$v(V)$}, audio-visual - \emph{$av(AV)$}, audio-only - \emph{$a(A\neg V)$} and video-only - \emph{$v(V\neg A)$} inputs. The embedding models CAV-MAE, DeepAVFusion, AV-Siam, and Equi-AV were finetuned on the \vggsound{} training set. We report metrics for $k=1$ here and for other $k$ in \cref{sec:additional-quantitative-analysis}. The closed-source multi-modal foundation models Gemini and open-sourced models use a zero-shot evaluation protocol and LLM-assisted protocol respectively. }\label{tab:main}
\vspace{-0.7em}
\end{table*}

\noindent We additionally measure the negative impact of using multi-modal inputs. In particular, $\mu$ is a new metric we propose to measure a model's \emph{modality confusion} ($\mu$).
We define it as
\begin{equation}\label{eq:mu}
    \mu_\text{M} = 100 \cdot \frac{\sum_{x \in M} \mathcal{I}[a(x) \text{-correct} \cap \, av(x)\text{-wrong}]}{N_{total}},
\end{equation}
where $M \in [A, V, A \cap V]$ and their associated modality inputs are $m \in [a, v]$, correct/wrong is determined as in the \emph{Hit} score (with $k=1$ for embedding models). $N_{total}$ refers to the total number of samples.
$\mu$ measures the fraction of samples a model correctly classified given an input modality but got wrong when using both modalities simultaneously.
We additionally report $\mu_{A \cap V}$ as the percentage of samples a model could solve \emph{in either modality} unimodally but could not solve multi-modally.
In other words, the modality confusion $\mu$ captures how frequently a model is distracted by an additional input modality, which can indicate the unsuccessful merging of modalities.

\takeaway{We propose a new metric, \emph{modality confusion} $\mu$, that measures how frequently a model is distracted by an additional input modality; see \cref{eq:mu}.}

\subsection{Re-evaluating the state of the art}\label{sec:reeval-sota}
We present the benchmark performance of state-of-the-art audio-visual models in \cref{tab:main}.

\mypara{Overall performance}
Unsurprisingly, all models perform best with access to both input modalities (\emph{AV}).
Across all metrics, both open- and closed-source general-purpose models perform comparably to the purpose-built embedding models CAV-MAE and AV-Siam. This indicates that foundation models have reached — and for some modalities exceeded — the performance of specialised models.
However, the  embedding models finetuned on \vggsound{} generally have stronger unimodal performance with audio inputs (\emph{A}) compared to visual inputs (\emph{V}).
Interestingly, this trend is reversed for most foundation models, which seem to be biased towards visual inputs, with unimodal video performance (\emph{V}) being substantially higher than unimodal audio performance (\emph{A}).

\takeaway{\emph{Foundation models} perform comparably to finetuned \emph{embedding models}. Embedding models more heavily rely on audio cues than on \emph{visual} ones, while foundation models exploit \emph{visible} cues rather than \emph{audible} ones, see \cref{tab:main}.}

\mypara{Modality confusion}
The \emph{modality confusion score} $\mu$ shows that, for all models, a notable fraction of test samples (4–11\%) were misclassified when an additional modality was included—despite being correctly classified with unimodal input. Furthermore, for all models, a small portion of test samples is not solvable multi-modally even though they were solvable in both modalities alone ($\mu_{A \cap V}$). In addition, for the majority of foundation models, $\mu_{A}$ is higher than $\mu_{V}$, indicating that these models forget audible labels more often than visible labels when a second modality is introduced. This suggests that, in such cases, they prefer visible information over audible information. Interestingly, this phenomenon is reversed for the majority of embedding models. This insight is made possible by \vggsounder{}'s per-label modality annotations and shows that all models are susceptible to being distracted given an additional modality. This is a concerning issue for multi-modal models since they should preserve unimodal capabilities when adding a second modality. Being able to evaluate this behaviour on the \vggsounder{} benchmark is a first step towards enabling the development of mitigation strategies, eventually resulting in stronger audio-visual models.

\takeaway{Our \emph{modality confusion score} reveals that \emph{all models} are negatively impacted by additional modalities for a substantial amount of samples (see \cref{tab:main}) and provides a means to quantify modality ensembles.}

\begin{figure}[ht]
    \centering
    \includegraphics[width=\linewidth, trim=0 30 0 0]{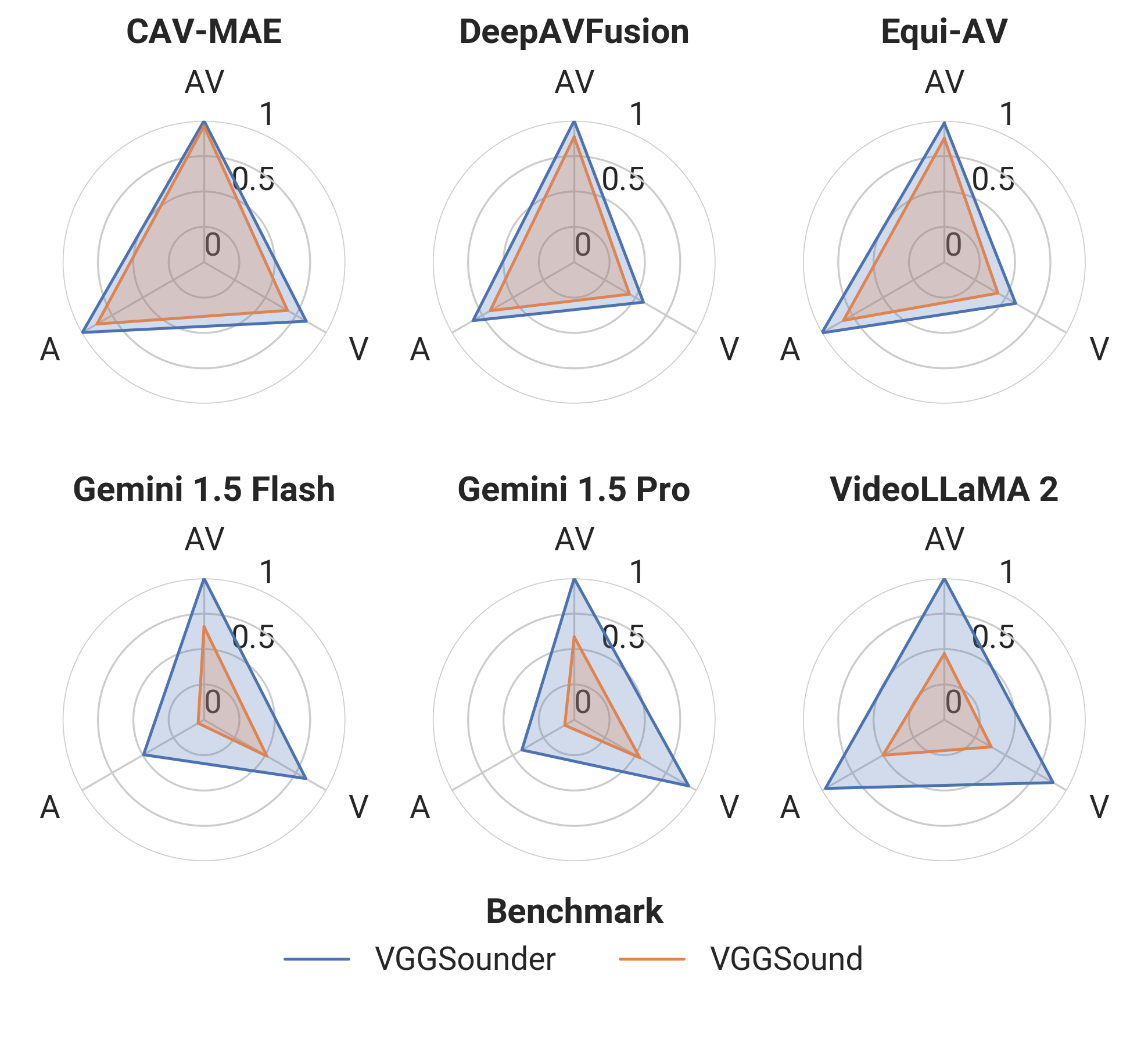}
    \caption{\textbf{\vggsounder{} more accurately captures model performance across input modalities}.
    We show the Hit score on \vggsounder{} and accuracy on \vggsound{}, normalised by the per-model maximum performance on each benchmark. Specifically for foundation models, we observe a significant difference in performance between \vggsound{} and \vggsounder{}.}
    \label{fig:rada}
    \vspace{-0.7em}
\end{figure}

\mypara{Performance across modalities}
\cref{fig:rada} shows the performance profiles across modalities.
At first glance, we can see that VideoLLaMA-2's performance is well balanced for different input modalities, while models from the Gemini family distinctly underperform on audio inputs. In contrast, embedding models exhibit a moderate balance across modalities, with DeepAVFusion and Equi-AV showing slight underperformance for visual input. 

\noindent As \cref{fig:rada} also illustrates, profiling of this kind is enabled through the modality annotations in \vggsounder{}. In contrast,
\vggsound{} assumed that all classes are perceivable in both modalities, and did not account for background sounds. This resulted in consistent under-evaluation of foundation models (that were not finetuned on \vggsound{}) for audio inputs.

\noindent In addition to the radar plot in \cref{fig:rada}, we provide results on \vggsound{} in \cref{sec:additional-quantitative-analysis}, showing that all models have substantially lower performance than their hit scores in \cref{tab:main}. This confirms that many model predictions were incorrectly flagged as false positives in \vggsound{} due to the incomplete ground-truth labels, painting a distorted picture of models' limitations.

\takeaway{\vggsounder{}'s more complete ground-truth labels allow for more accurate, modality-specific profiling of model performance (see \cref{fig:rada}).}

\subsection{Performance analysis using meta-classes}\label{sec:meta-classes}
\begin{table*}[t]
\centering
\resizebox{\textwidth}{!}{
\begin{tabular}{l S[table-format=-2.2] S[table-format=-2.2] S[table-format=-2.2] S[table-format=-2.2] S[table-format=-2.2] S[table-format=-2.2] S[table-format=-2.2] S[table-format=-2.2] S[table-format=-2.2] S[table-format=-2.2] S[table-format=-2.2] S[table-format=-2.2] S[table-format=-2.2]}
\toprule
& \multicolumn{3}{c}{\classname{background music}} & {\classname{voice over}} & \multicolumn{6}{c}{\classname{static image(s)}} & \multicolumn{3}{c}{neither} \\
\cmidrule(lr){6-11}
& \multicolumn{3}{c}{$\Delta$ \textbf{$\mathbf{F_1}$}} & $\Delta$ \textbf{$\mathbf{F_1}$} & \multicolumn{2}{c}{$\Delta$ \textbf{$\mathbf{F_1}$}}  & \multicolumn{2}{c}{\textbf{Sub. Acc.} w/}  & \multicolumn{2}{c}{\textbf{Sub. Acc.} w/o} & \multicolumn{3}{c}{$\Delta$ \textbf{$\mathbf{F_1}$}}\\
\cmidrule(lr){2-4} \cmidrule(lr){5-5} \cmidrule(lr){6-7} \cmidrule(lr){8-9} \cmidrule(lr){10-11} \cmidrule(lr){12-14}
\textbf{Model} & \emph{a} & \emph{v} & \emph{av} & \emph{a} & \emph{a} & \emph{v} & \emph{a} & \emph{v} & \emph{a} & \emph{v} & \emph{a} & \emph{v} & \emph{av} \\

\midrule[0.75pt]
\multicolumn{14}{l}{\emph{Embedding Models}}\\
\addlinespace[2pt] 

CAV-MAE & -6.80 & -2.27 & -3.45 & -11.93 & 4.93 & -5.04 & 25.31 & 26.75 & 15.68 & 24.09 & -3.42 & -1.20 & -1.28 \\
DeepAVFusion & -7.38 & -3.64 & -5.44 & -12.09 & 3.29 & -5.45 & 18.43 & 15.40 & 12.07 & 13.46 & -3.98 & -0.70 & -2.36 \\
Equi-AV & -7.58 & -2.85 & -5.11 & -10.72 & 5.12 & -7.42 & 22.46 & 11.25 & 13.64 & 13.57 & -3.41 & -1.08 & -2.24 \\
AV-Siam & -7.16 & -4.60 & -6.00 & -12.46 & 4.55 & -9.69 & 24.50 & 23.71 & 15.12 & 24.70 & -3.70 & -1.71 & -1.87 \\

\midrule
\multicolumn{14}{l}{\emph{Closed-source Foundation Models}}\\
\addlinespace[2pt] 

Gemini 1.5 Flash & 1.38 & -3.45 & -4.35 & 23.73 & -4.47 & -2.18 & 2.11 & 19.60 & 2.44 & 18.40 & 4.77 & -1.35 & -1.67 \\
Gemini 1.5 Pro & 1.50 & -0.85 & -0.82 & 23.73 & -4.48 & -0.75 & 3.35 & 32.83 & 3.63 & 27.02 & 5.41 & -0.97 & -1.74 \\
Gemini 2.0 Flash & -0.78 & -2.70 & -4.19 & -1.88 & 1.35 & -2.67 & 4.34 & 18.69 & 1.93 & 15.88 & -0.57 & -0.84 & -1.34 \\

\midrule
\multicolumn{14}{l}{\emph{Open-source Foundation Models}}\\
\addlinespace[2pt] 

VideoLLaMA 2 & -4.17 & -2.34 & -2.04 & -7.62 & 4.61 & -3.02 & 21.77 & 25.99 & 14.69 & 23.66 & -2.04 & -0.54 & -0.45 \\
Unified-IO 2 & -9.88 & 3.12 & -2.42 & -3.45 & 1.63 & -3.41 & 20.29 & 14.89 & 13.29 & 15.02 & -2.25 & 0.44 & -0.90 \\
PandaGPT & -4.13 & 1.18 & -0.31 & 7.27 & -6.33 & -3.02 & 2.30 & 6.99 & 3.33 & 5.47 & 0.58 & 0.25 & -0.06 \\
OLA & -8.58 & 3.10 & 2.59 & 15.49 & -6.84 & -7.34 & 16.87 & 8.66 & 14.80 & 10.95 & 1.96 & 0.04 & 0.12 \\    

\bottomrule
\end{tabular}
}
\caption{
    \textbf{Summary of performance differences in the presence/absence of meta-classes.}
{Difference in $\mathbf{F_1}$ scores ($\Delta$ \textbf{$\mathbf{F_1}$}})  for audio-visual video classification on \vggsounder{} between videos \emph{with} a meta-class and those \emph{without} it. Positive numbers ($\Delta$) indicate better performance when the meta-class is present.
Additional results are provided in \cref{sec:additional-quantitative-analysis}.
}\label{tab:meta-classes}
\vspace{-0.7em}
\end{table*}

\vggsounder{} includes annotations of three meta-classes for each sample:
\classname{background music} indicates whether samples contain music without a visible source, 
\classname{voice over} similarly marks speech without a visible source, and 
\classname{static image(s)} flags that the visual stream consists of one or only a few static visual frames.

\noindent These new meta-classes allow us to evaluate the model behaviour in challenging scenarios where information from one modality dominates.
We consider the performance difference between samples that \emph{do} contain a meta-label and samples that \emph{do not} contain it.
Positive numbers indicate that the models perform better on the subset with meta-labels.
In \cref{tab:meta-classes}, we summarise the main findings, focussing on $F_1$-score as the most balanced multi-label metric. Additional results are provided in \cref{sec:additional-quantitative-analysis}.

\mypara{Background music}
All models perform worse on video samples containing background music.
This indicates that it is challenging to decouple background audio from the rest of the video. The evaluated models are not good at differentiating between sound sources without visual cues, e.g.\ predicting different instruments in the background music. 

\mypara{Voice over}
In contrast to background music, we observe a clear difference between embedding models and foundation models for samples with voice-over.
While the audio classification performance of embedding models drops substantially. This drop is only slight for VideoLLaMA-2 and Unified-IO 2, and the performance of other foundation models even improves.
This indicates that the foundation models are less distracted by voice-over.

\mypara{Static image(s)}
The impact of static images is more nuanced:
First, audio classification performance improves for embedding models while it decreases for half of the  foundation models.
This shows that the purpose-built, \vggsound{}-finetuned embedding models can more accurately predict specific sounds in the absence of other cues.

\noindent Second, visual classification performance on static images drops for all models, suggesting that models rely on rich temporal cues to make accurate predictions.

\mypara{Samples without any meta-label}
When comparing the model performance on samples without background music, static images, or voice-over annotations (column \emph{neither} in \cref{tab:meta-classes}) to the performance on samples that contain either of meta classes, we see a performance gain.
This finding concludes that these three categories form challenging subsets of the dataset.

\takeaway{Samples with \emph{background music}, \emph{static images(s)}, and \emph{voice over} provide distinct challenges for each model (see \cref{tab:meta-classes}). This highlights \vggsounder{}'s value for comprehensive model evaluation.}

\subsection{Impact of \vggsounder{} labels}\label{sec:ablations}
\begin{table}[t]
\centering
\resizebox{\linewidth}{!}{
\begin{tabular}{l S[table-format=-2.2] S[table-format=-2.2] S[table-format=-2.2] S[table-format=-2.2] S[table-format=-2.2] S[table-format=-2.2]}
\toprule
& \multicolumn{3}{c}{Human labels \small$\uparrow$} & \multicolumn{3}{c}{Auto labels \small$\uparrow$} \\
\cmidrule(lr){2-4} \cmidrule(lr){5-7}
\textbf{Model} & \emph{A} & \emph{V} & \emph{AV} & \emph{A} & \emph{V} & \emph{AV} \\
\midrule
Gemini 1.5 Flash & 29.28 & 14.59 & 16.36 & 0.48 & 0.93 & 1.51 \\
Gemini 1.5 Pro & 28.61 & 25.52 & 27.63 & 0.31 & 1.99 & 2.10 \\
Gemini 2.0 Flash & 8.80 & 12.16 & 11.13 & 0.22 & 1.12 & 1.28 \\
\bottomrule
\end{tabular}
}
\caption{\textbf{Impact of added labels using different strategies in \vggsounder{}.} We show the change in multi-label classification performance ($\Delta$ Hit) when adding automatically added (Auto) or human-annotated (Human) labels to \vggsounder{}, and compare to the original \vggsound{} data.}
\label{tab:ablate-heuristic}
\vspace{-0.7em}
\end{table}

\noindent
Our relabelling pipeline adds two types of labels to those in \vggsound{}: (1) automatically generated labels based on synonymous classes and sub-/superclasses, and (2) human-curated labels.
In \cref{tab:ablate-heuristic} , we ablate the impact of each type of added labels in terms of the relative performance gains (Hit score).
A complete breakdown of the effects across all models and metrics is provided in \cref{sec:additional-quantitative-analysis}.
While performance is consistently higher with automatically added labels (Auto), the increase is noticeably smaller than that for human-curated labels.
Paired with the observation that models do frequently predict correct classes that were not part of the original \vggsound{} label set, this indicates that human-curated labels better cover the ground truth.

\takeaway{Automatically added labels are an important step, but human-curated labels have a bigger effect on eliminating incorrectly flagged false positives, underscoring the value of accurate human annotation.}

\section{Discussion} 

\mypara{Choice of \vggsound{} as base dataset} \vggsound{} is commonly used to evaluate audio-visual models on the multi-modal video classification task. As it is currently the most suitable testbed for audio-visual classification tasks (due to its size, diversity of categories, non-constrained setting, and relatively strong audio-visual correspondence), it serves as an optimal starting point for our substantially improved \vggsounder{} benchmark with a multi-label evaluation protocol for foundation models that makes the benchmark suitable for meaningful evaluation.

\mypara{Multi-label vs single-label classification} Video content is inherently complex, often containing multiple co-occurring objects and actions both within and across modalities. This makes it unlikely that a given clip belongs to just one class. Therefore, our \vggsounder{} extends the \vggsound{} test set to multi-label classification. Unlike models trained on a narrow single-label dataset, foundation models develop versatile representations.

\section{Conclusion}
We introduced a modality-aware evaluation set for audio-visual foundation models. \vggsounder{} builds on the widely used \vggsound{} dataset by adding: (a) comprehensive human annotations for missing classes, (b) specifying modality information per label, (c) introducing specialised meta-labels for frequently occurring real-world challenges, and (d) using heuristic methods to improve label quality. 

\noindent Through our newly introduced metric, modality confusion, we observe that incorporating additional modalities does not necessarily yield better results. Models often become more confused on a substantial subset of test samples. 
Furthermore, finetuned embedding models tend to rely heavily on audio cues, while foundation models depend more on visual information. Additionally, our meta-label analysis highlights distinct challenges across various specialised yet commonly occurring scenarios such as background music, static images, and voice-overs. Overall, we hope that the \vggsounder{} benchmark will advance the evaluation and development of foundational audio-visual models.

\newpage
\section*{Acknowledgements} 
The authors would like to thank Felix Förster, Sayak Mallick, Amirshayan Nasirimajd, Jianzhe Liu, Danqing Chen, Noah Latzel, Björn Kremser, Simon Bleßmann and Prasanna Mayilvahananan for their help with data annotation, Felix Förster and Monica Riedler for proofreading the paper, and Thomas Klein and Shyamgopal Karthik for their help in setting up MTurk. They also thank numerous MTurk workers for labelling.

\noindent This work was in part supported by the BMBF (FKZ: 01IS24060, 01IS24085B, 01IS18039A), the DFG (SFB 1233, project number: 276693517), and the Open Philanthropy Foundation funded by the
Good Ventures Foundation. WB acknowledges financial support via an Emmy Noether Grant funded by the German Research Foundation (DFG) under grant no. BR 6382/1-1. WB is a member of the Machine Learning
Cluster of Excellence, EXC number 2064/1 – Project number 390727645. This research utilised compute resources at the Tübingen Machine Learning Cloud, DFG FKZ INST 37/1057-1 FUGG.
The authors thank the IMPRS-IS for supporting TW.

\paragraph{Author contributions} 
This project was co-led and coordinated by ASK, DZ, TW, and AP.
TW, DZ, and ASK analysed and identified issues with \vggsound{}.
TW and DZ implemented the annotation pipeline with input from ASK, AP, and WB.
TW ran annotations on MTurk and produced the final label set with support from DZ and input from ASK, AP, and WB.
DZ implemented the in-house annotation pipelines with input from ASK, TW, and AP.
DZ trained in-house versions of multiple models with help from ASK.
DZ evaluated all models with support from ASK and input from TW and AP.
TW, ASK, DZ, and AP wrote the manuscript with input from WB and MB. 
TW and DZ created the figures with feedback from ASK, AP, and WB.
MB provided helpful feedback throughout the project.

{
    \small
    \bibliographystyle{ieeenat_fullname}
    \bibliography{main}
}

\clearpage
\newpage

\appendix
\onecolumn
\begin{center}
{\Large \bf Supplementary Material:\\ \vggsounder{}: Audio-Visual Evaluations for Foundation Models \par}
\vspace*{24pt}
\end{center}

\section{\vggsounder{}: Relabelling \vggsound{}}
\label{sec:vggsounder-relabelling-vggsound}
In the main paper, we highlighted several critical shortcomings of \vggsound{}, such as co-occurring classes, partially overlapping class definitions, multiple classes per sample, and modality misalignment. This appendix provides additional details about the relabelling process for obtaining the \vggsounder{} benchmark, addressing the specific issues identified in \vggsound{}.

\subsection{Labelling of the gold-standard subset}
\begin{figure}[ht]
    \centering
    \includegraphics[width=\textwidth, trim=10 40 10 10]{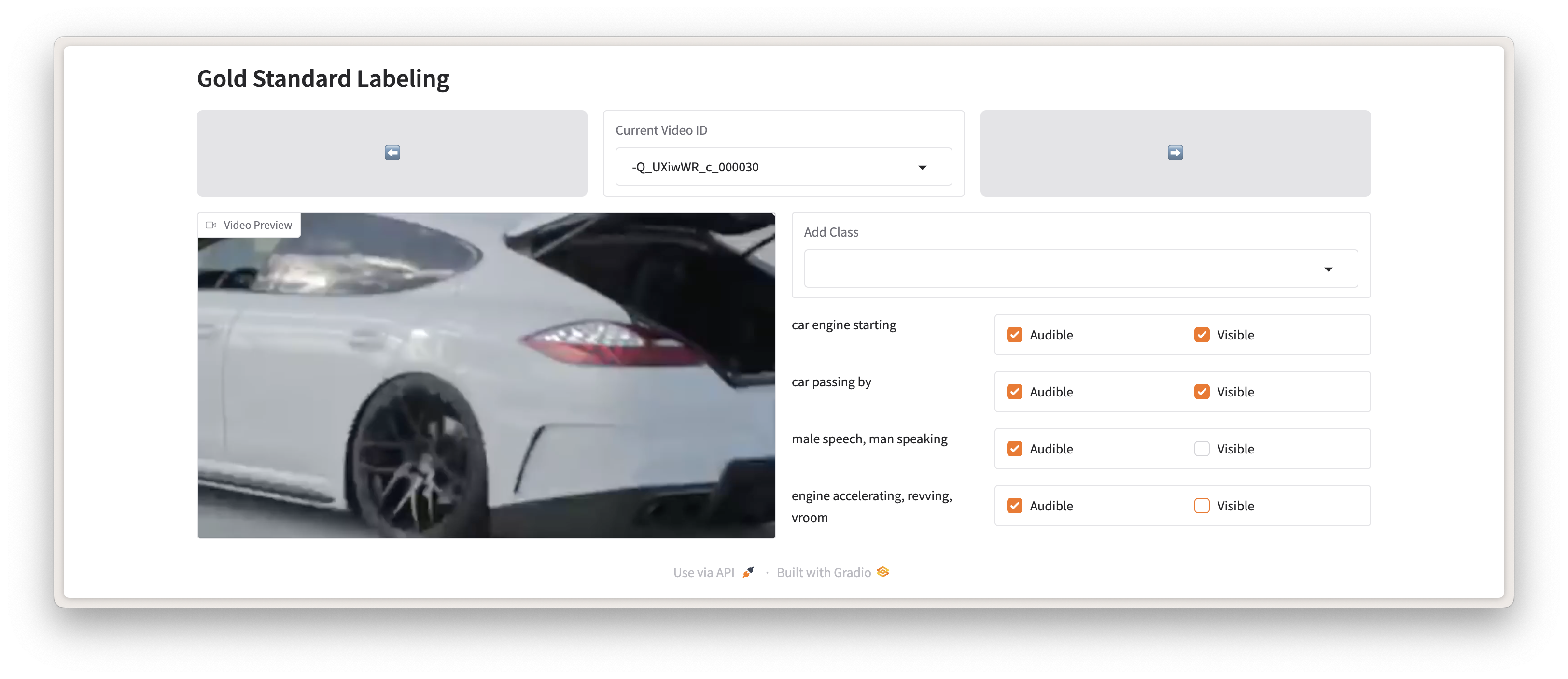}
    \caption{Interface used to annotate the gold standard set in-house.}
    \label{fig:interface-goldstandard}
\end{figure}

As described in \cref{sec:vggsounder}, we started by creating a high-quality reference subset (gold-standard) for reliable label verification. Four experienced computer vision researchers manually annotated randomly selected 10-second videos from the \vggsound{} test set. Annotators labelled classes clearly present either audibly, visually, or both. We ensured full class coverage by continuing the annotation process until all classes appeared at least once, resulting in 417 samples. These annotations were merged using majority voting. The annotation interface employed in this phase is illustrated in \cref{fig:interface-goldstandard}.

\noindent The annotators were instructed to try to identify all audible and visible classes in the video, including hard cases when background music contains several instruments that compose the melody. For instance, a common instrument for the country music genre would be \classname{playing the drum kit}, \classname{female singing}, \classname{male singing}, \classname{playing the bass guitar},  \classname{playing electric guitar} etc. The annotators are expected to do their best to identify all of the instruments. 

\callout{
Gold-standard samples serve as high-quality annotations for further labellers' cross-validation and automatic quality assessment. If a labeller shows a high agreement score with the gold-standard labels, we expect them to have high-quality labels outside of the gold-standard subset.
}

\noindent While analysing gold-standard labels, we made several interesting observations (see \cref{tab:goldstandard-stats}): 

\begin{enumerate}
    \item There is a significant portion of samples in the gold-standard set for which the original \vggsound{} labels (24.46\%) are absent.
    \item  The proportion of classes that are only audible across all samples is significantly higher than that of the visible ones.
\end{enumerate}

\newpage

\begin{table}[ht]
\centering
\begin{tabular}{lr}
\hline
\textbf{Metric} & \textbf{Value} \\
\hline
Samples & 417 \\
Original class correct & 283 (67.87\%) \\
Original class audible & 39 (9.35\%) \\
Original class visible & 22 (5.88\%) \\
\hline
Original class absent & 102 (24.46\%) \\
\hline
Original class is only class & 71 (17.03\%) \\
Classes total & 309 \\
\hline
Classes only visible & 6 \\
Classes only audible & 25 \\
\hline
Average labels added per sample & 1.39 \\
\hline
\end{tabular}
\caption{Relabelling statistics for the gold-standard subset.}
\label{tab:goldstandard-stats}
\end{table}

\noindent While we cannot fix the second issue without substituting the dataset, the first issue quantifies the error introduced by \vggsound{} and its automatic labelling and verification and can be eliminated with human labelling. 

\vspace{0.7\baselineskip}

\begin{figure}[ht]
    \centering
    \includegraphics[width=\textwidth, trim=10 40 10 10]{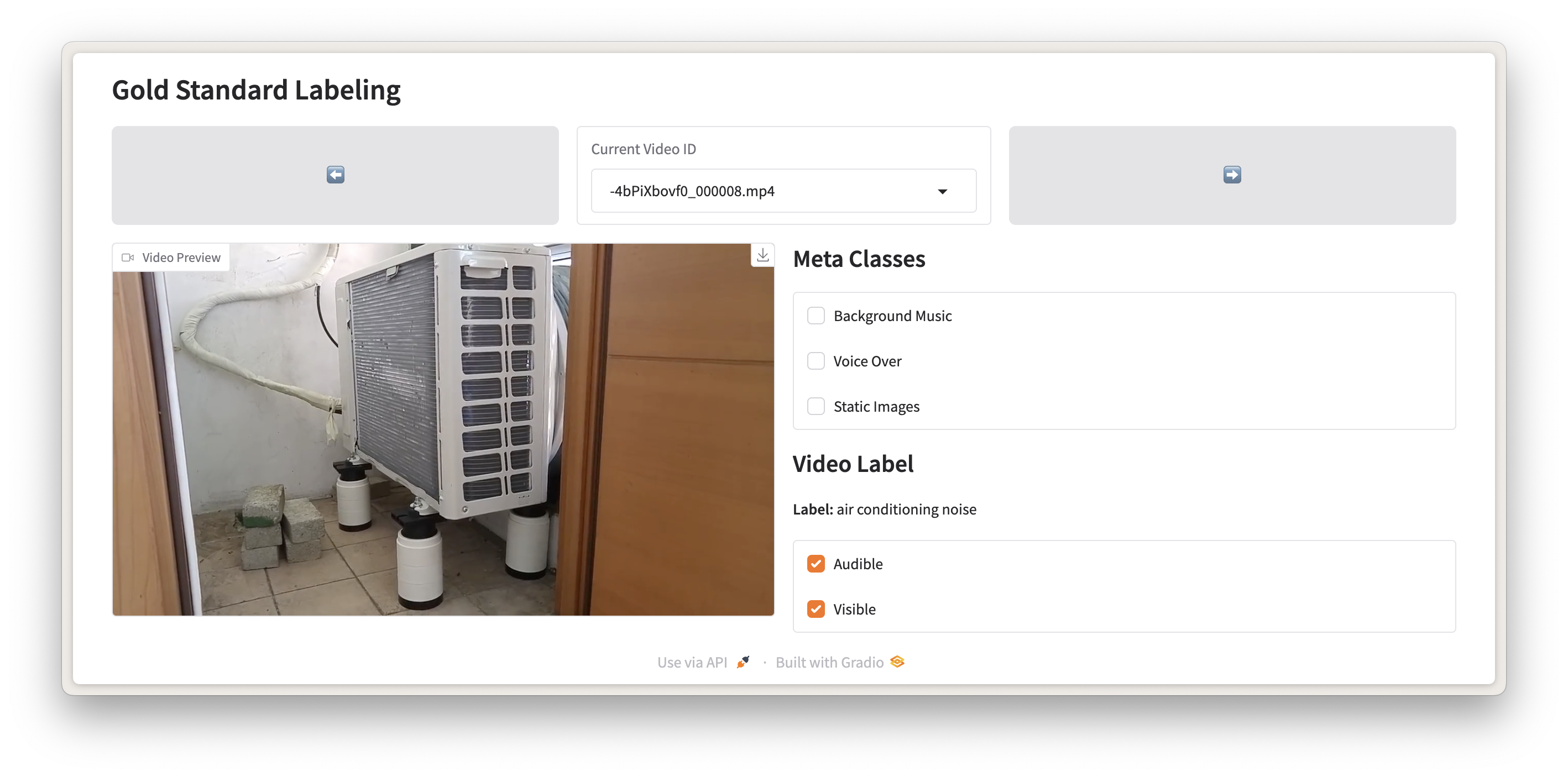}
    \caption{Interface used in-house to annotate the original labels in the \vggsound{} test set.}
    \label{fig:felix-interface-goldstandard}
\end{figure}

\noindent We ran a second round of gold-standard annotations where one computer vision expert checked all 15446 samples and annotations in the \vggsound{} test set for their validity and enriched the correct labels with modality annotations. The interface for this annotation is illustrated in \cref{fig:felix-interface-goldstandard}.

\noindent The second set of gold-standard labels firstly enriched the original \vggsound{} labels with modality annotations, but most importantly confirmed and further improved the estimates in \cref{tab:goldstandard-stats} resulting in the following observation: 
\callout{Around 48.43\% of the original \vggsound{} test samples
have either incorrect target labels or misaligned modalities.}

\noindent The two sets of gold-standard annotations, while having mixed reliability (cross-validation with four people vs.\ one person), serve as a strong grounding signal for our subsequent MTurk annotation pipelines. 

\subsection{Label proposals}
\label{sec:label-proposal-app}
To effectively scale human annotations to the entire test set, and to simplify the job for MTurk annotators, we introduced a label proposal generation strategy that combines state-of-the-art audio-visual model predictions with label heuristics. We considered the following steps in our label proposals:

\begin{enumerate}
    \item  \mypara{Model predictions:}
    \begin{itemize}
        \item We provide the original \vggsound{} label, extended with modality annotations curated by an in-house labeler, as well as the top-1 predictions of the following models \footnote{Gemini 2.0 Flash, VideoLLaMA-2, Unified‑IO-2, Panda-GPT, and Ola were not used when the proposals were generated.} with visual and audio-visual inputs:
        \begin{itemize}
            \item CAV-MAE
            \item AV-Siam 
            \item Equi-AV 
            \item DeepAVFusion
            \item Gemini 1.5 Flash
            \item Gemini 1.5 Pro
        \end{itemize}
        \item We further included the top-5 predictions when using audio inputs from the same models.
    \end{itemize}

\item \mypara{Consensus labels:}
\begin{itemize}
    \item We created a secondary pool from the top-10 predictions across all modalities from AV-Siam, CAV-MAE, and Equi-AV. Additionally, labels associated with the highest 60,000 logits or probabilities across the dataset were added.
    \item Labels were proposed from this pool if at least two models independently agreed on their presence.
\end{itemize}

\item \mypara{Common classes:}
\begin{itemize}
    \item Regardless of model predictions, we always proposed frequently occurring classes such as: 
    \begin{center}
        \classname{wind noise}, \classname{wind rustling leaves}, \classname{male speech, man speaking}, \classname{female speech, woman speaking}, \classname{child speech, kid speaking}, \classname{bird chirping, tweeting}, \classname{cricket chirping}, \classname{sea waves}.
    \end{center}
\end{itemize}

\end{enumerate}

\noindent This strategic combination ensured an average of 30 proposals per video, achieving approximately 93\% recall relative to the gold-standard set annotations.

\subsection{Annotation pipeline}\label{sec:app-final-pipeline}
Our annotation pipeline consists of the following consecutive stages: 

\begin{center}
\begin{tikzpicture}[
  node distance=1.1cm,
  box/.style={
    draw,
    rectangle,
    minimum height=9mm,
    minimum width=2.8cm,
    align=center
  },
  arrow/.style={->, thick}
]

\node[box] (mturk) {
  MTurk Annotations\
};

\node[box, right=of mturk] (inhouse) {
  In-House Annotations
};

\node[box, right=of inhouse] (lvm) {
  LVM Proposals
};

\node[box, right=of lvm] (classes) {
  Final labels
};

\draw[arrow] (mturk) -- (inhouse);
\draw[arrow] (inhouse) -- (lvm);
\draw[arrow] (lvm) -- (classes);

\end{tikzpicture}
\end{center}

\noindent This pipeline produces the cleanest final label set that we refer as \vggsounder{}, and achieves 0.68 macro-averaged F1 on the gold-standard data. 

\noindent \\ As noted in the main paper \cref{sec:experiments}, our default evaluation subset excludes samples with \classname{background music} label; however, we provide \vggsounder{} with all samples and allow the user to choose their preferred regime, if necessary.

\subsection{MTurk Annotations}
\label{sec:human-lalbel-app}
\begin{figure}[t]
    \centering
    \begin{subfigure}[t]{0.72\linewidth}
        \centering
        \includegraphics[width=\linewidth]{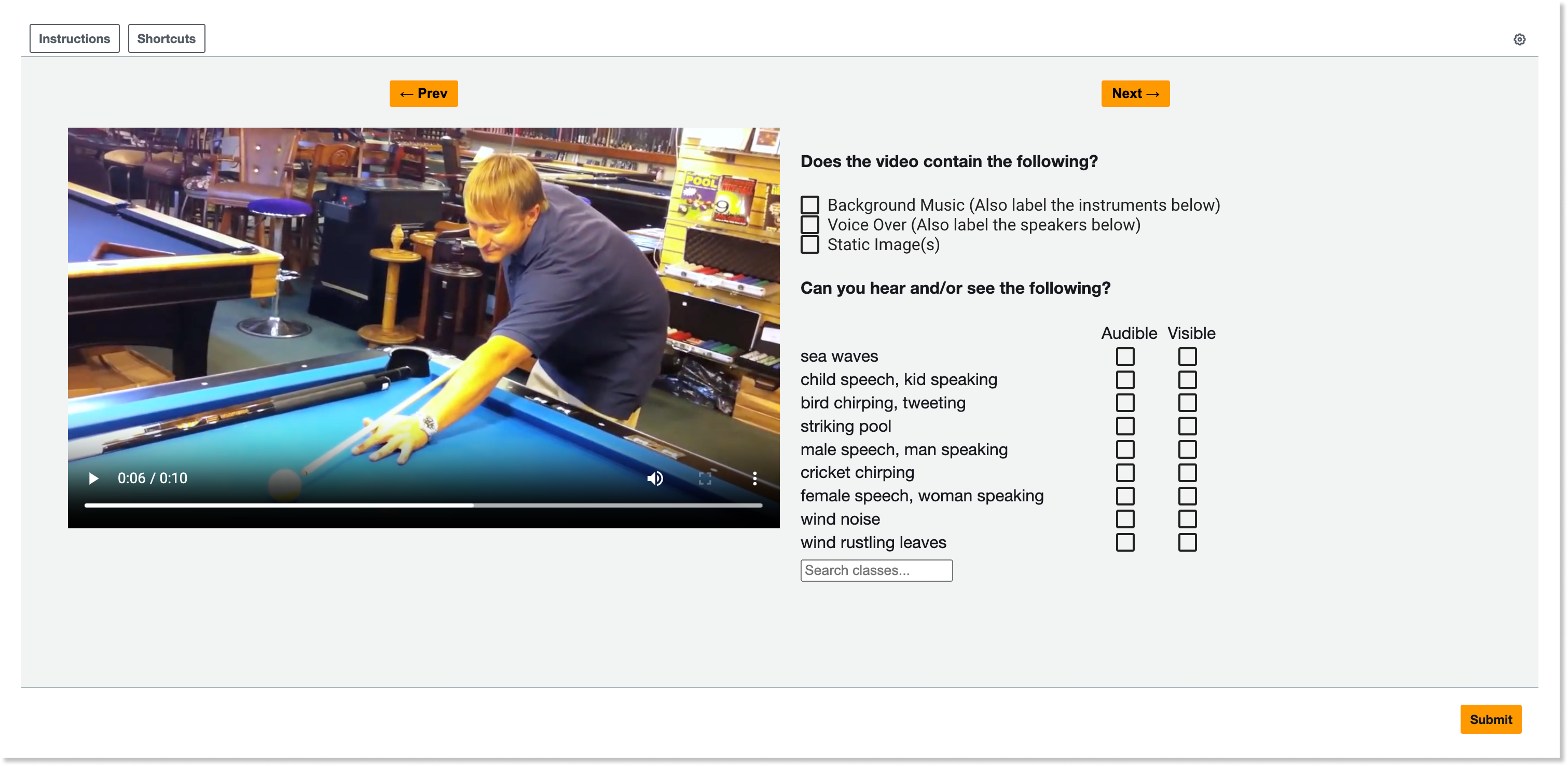}
        \caption{\textbf{Example labelling interface for one video sample.}}
        \label{fig:mturk_onesample}
    \end{subfigure}%
    \hfill
    \begin{subfigure}[t]{0.25\linewidth}
        \centering
        \includegraphics[width=\linewidth]{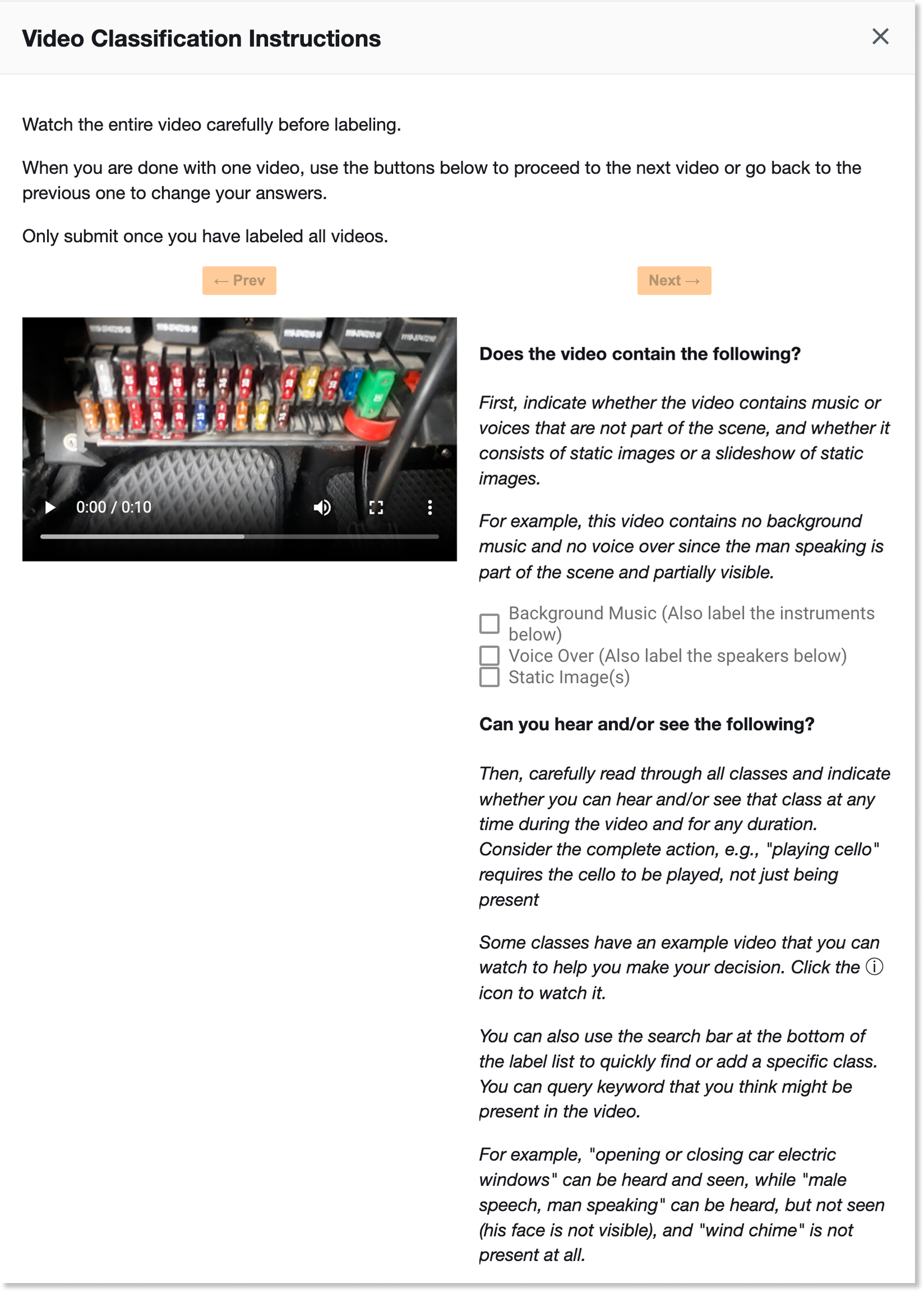}
        \caption{\textbf{Labelling instructions}}
        \label{fig:mturk_instructions}
    \end{subfigure}
    \caption{\textbf{Labelling interface and instructions for our full annotation pipeline that we ran on MTurk.} (a) Crowd workers are presented with a 10-second long video clip from the \vggsound{} test set, along with label proposals. They are tasked to select if those or additional \vggsound{} classes are audible or visible in the video clip. Furthermore, the workers are asked about meta-classes, such as background music, voice-over, and static images. They also have the option of searching for new classes that are missing in the proposals. (b) Labelling instructions provided to workers on Amazon Mechanical Turk before labelling the first video sample.}
    \label{fig:mturk_interface}
\end{figure}

Following our proposal strategy \cref{sec:label-proposal-app}, we conducted extensive human annotation via Amazon Mechanical Turk (MTurk) to verify and expand the automatically generated proposals:
\begin{itemize}
    \item \mypara{Worker qualifications:} We used two types of Amazon Mechanical Turk (AMT) worker qualifications for our annotations. Half of the tasks were completed by AMT Masters with approval rates above 98\%, while the other half were assigned to non-Master workers who also maintained approval rates above 98\%. We adopted this approach as the larger non-Master worker pool resulted in significantly faster task completion.
    \item \mypara{Annotation interface:} Annotators reviewed each video to confirm the presence and modality (audible, visible, or both) of proposed labels. They could also suggest additional missing labels. Workers received detailed instructions (see \cref{fig:mturk_instructions}), and the annotation interface used is presented in \cref{fig:mturk_onesample}.
    \item \mypara{Quality control:} Videos were grouped into batches of 20, each containing two gold-standard samples as catch trials. Batches scoring below 25\% F1-score on these catch trials were rejected and reassigned.
\end{itemize}

\noindent \\ We collect all crowd-sourced (Mechanical Turk) and some in-house annotations collected prior to MTurk labelling, so that every video is reviewed by at least 3 annotators.  Annotators are ranked by their mean accuracy score on the gold-standard clips, and the top three for each video are retained. Their votes are then combined with the Dawid-Skene algorithm.

\subsubsection{Dawid--Skene merging details}
\label{sec:dawid-skene}
An annotation signal is represented as a triple (sample, class, modality) and is treated as a binary task: whether that label is present in that modality. As stated above, we aggregate the per-annotator votes with the Dawid--Skene model~\cite{sinha2018fastdawidskenefastvote}, which jointly estimates, via expectation--maximisation (EM), a per-annotator confusion matrix and the latent true label of every task, so that systematically unreliable annotators are automatically down-weighted instead of counting equally. We seed the estimation in two ways. First, annotations that fall in our trusted gold-standard subset (five independent annotations each) are anchored to their known labels, fixing the latent variables for those tasks and propagating reliable structure to the rest. Second, each annotator's confusion matrix is initialised from their empirical error counts measured against the gold standard, giving the EM a calibrated starting point. Our expert in-house annotations carry no crowd history, so we add a strong diagonal pseudo-count prior ($50$ counts) that lets the model trust them from the first iteration. 

\subsection{In-House Annotations}
\begin{figure}[h!]
    \centering
    \includegraphics[width=\linewidth]{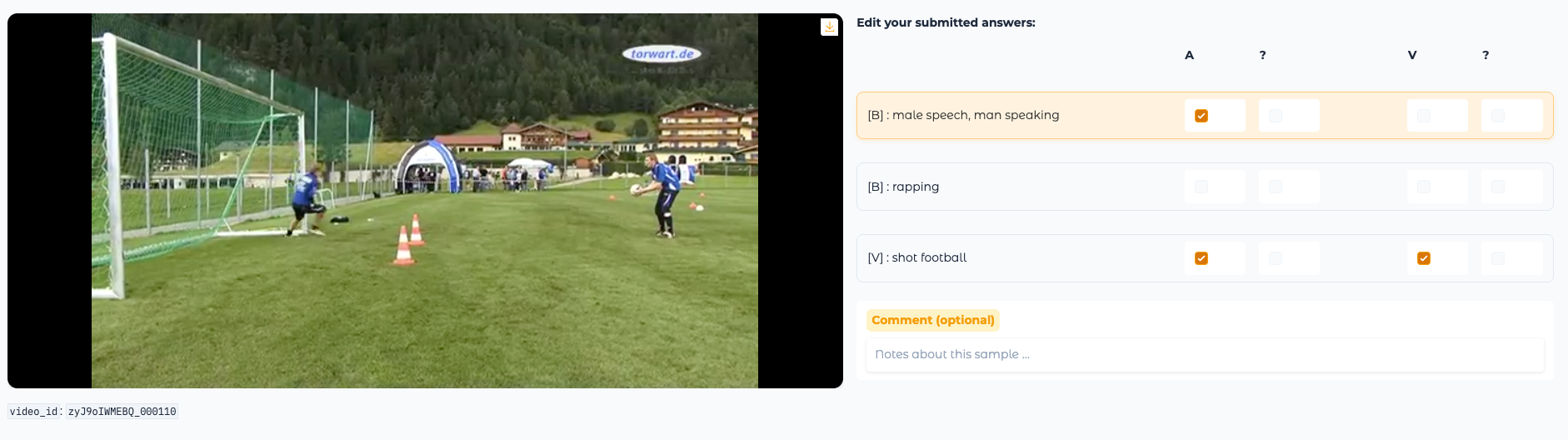}
    \caption{\textbf{Interface of the Gradio application.} That was used for the correction of labels for automatically flagged 3,616 videos. Annotators were present only with labels submitted by high-scoring MTurk labellers.}
    \label{fig:mturk_fix_interface}
\end{figure}

Post MTurk labelling, we additionally collected modality annotations internally for more than 3600 samples with the help of several reliable volunteers, all computer science experts with diverse backgrounds. They were presented with a similar interface (see \cref{fig:mturk_fix_interface}) and asked to correct the collected MTurk annotations, which were merged using majority voting. This was done only for a subset of 3,616 videos after automatic flagging, as these samples contained many incorrect labels despite being proposed by MTurk workers. In-house annotations are then applied over (override) MTurk labels. 

\subsection{LVM proposals} \label{sec:lvm-proposals-app}
We noticed that Dawid-Skene over-deletes some commonly appearing classes, such as \classname{wind noise}, \classname{wind rustling leaves}, \classname{male speech, man speaking}, \classname{female speech, woman speaking}, and decided to automatically compensate for it. We promoted the \texttt{gemini-3.5-flash} model, with label differences between datasets obtained via Dawid-Skene merging and majority voting merging, both computed on MTurk raw annotations. We then asked the LVM whether these classes are only audible or visible and genuinely present in the corresponding modality. We used the following prompt:

\begin{lstlisting}
Audit labels for a 10-second video clip that has BOTH a visual and an audio track.

A label EXISTS in AUDIO if and only if you can HEAR it AND the source genuinely occurs in the clip -- regardless of whether it is visible. (A bird chirping off-screen still exists in audio.) If you can only SEE the source but cannot hear it, it does NOT exist in audio.

A label EXISTS in VISUAL if and only if you can SEE the source genuinely present in the scene -- regardless of whether it is audible. If you can only HEAR it but cannot see it, it does NOT exist in visual.

For modality "A": exists = true only if it exists in audio (per the rule above).
For modality "V": exists = true only if it exists in visual.
For modality "AV": exists = true only if it exists in BOTH audio and visual.

For each label below, decide whether it exists in its stated modality: {labels_block}
Return valid JSON exactly:
{{
  "verifications": [
    {{
      "label": "...",
      "modality": "A",
      "exists": true,
      "confidence": 0.0,
      "reason": "short explanation"
    }}
  ]
}}
\end{lstlisting}

\noindent After a thorough check, we accepted all Gemini proposals. 

\begin{wraptable}[13]{r}{0.4\textwidth}
    \centering
    \vspace{-12em} 
    \resizebox{0.37\textwidth}{!}{
        \begin{tabular}{ll}
            \toprule
            \textbf{Class} & \textbf{Added Class} \\
            \midrule
            timpani & tympani \\
            tympani & timpani \\
            dog barking & dog bow-wow \\
            dog bow-wow & dog barking \\
            Barn swallow calling & Bird chirping, tweeting \\
            Eagle screaming & Bird squawking \\
            Canary calling & Bird chirping, tweeting \\
            Mynah bird singing & Bird chirping, tweeting \\
            Magpie calling & Bird squawking \\
            Warbler chirping & Bird chirping, tweeting \\
            Wood thrush calling & Bird chirping, tweeting \\
            Goose honking & Bird squawking \\
            Duck quacking & Bird squawking \\
            Penguins braying & Bird squawking \\
            Baltimore oriole calling & Bird chirping, tweeting \\
            Crow cawing & Bird squawking \\
            Airplane flyby & Airplane \\
            Baby babbling & People babbling \\
            Bull bellowing & Cattle mooing \\
            Cow lowing & Cattle mooing \\
            People eating noodle & People eating \\
            People eating apple & People eating \\
            Eating with cutlery & People eating \\
            Bathroom ventilation fan running & Running electric fan \\
            Striking bowling & Bowling impact \\
            \bottomrule
        \end{tabular}
    }
    \caption{Class mapping heuristics used to add synonymous classes and superclasses.}
    \label{tab:class-mapping}
    \vspace{-2em}
\end{wraptable}

\subsection{Final labels that include class heuristics}
To resolve ambiguous classes (see \cref{sec:vggsound-issues}), we included synonymous classes and related superclasses. For instance, \classname{cow lowing} led to including the superclass \classname{cattle mooing}. \cref{tab:class-mapping} summarises the class heuristics used.

\section{Class label frequency in \vggsounder{}}

\label{sec:class-label-frequency-in-vggsounder}

\begin{figure}[t!]
    \centering
    \begin{subfigure}[t]{0.32\textwidth}
        \centering
        \includegraphics[width=\linewidth]{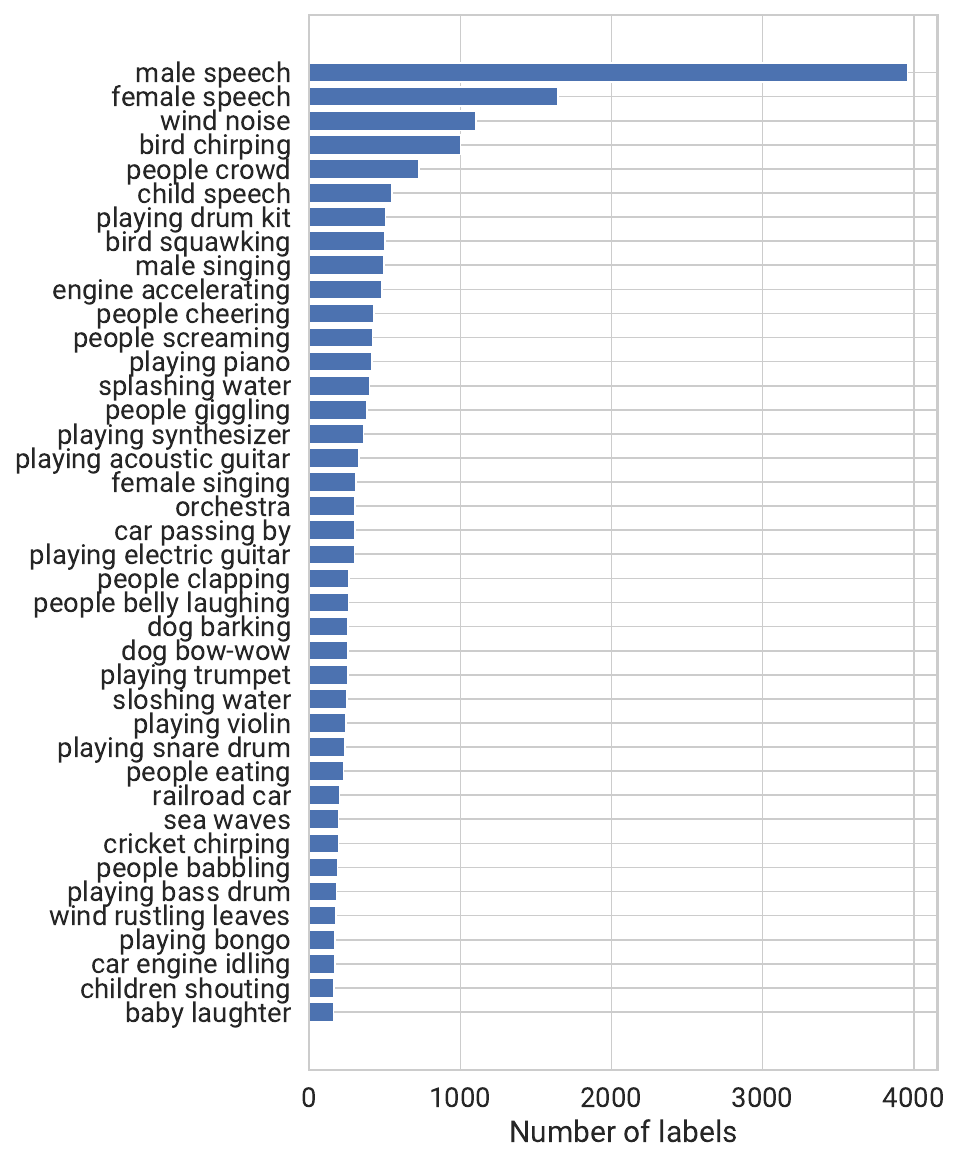}
        \caption{\textit{Audible} classes.}
        \label{fig:audio_freq}
    \end{subfigure}
    \hfill
    \begin{subfigure}[t]{0.32\textwidth}
        \centering
        \includegraphics[width=\linewidth]{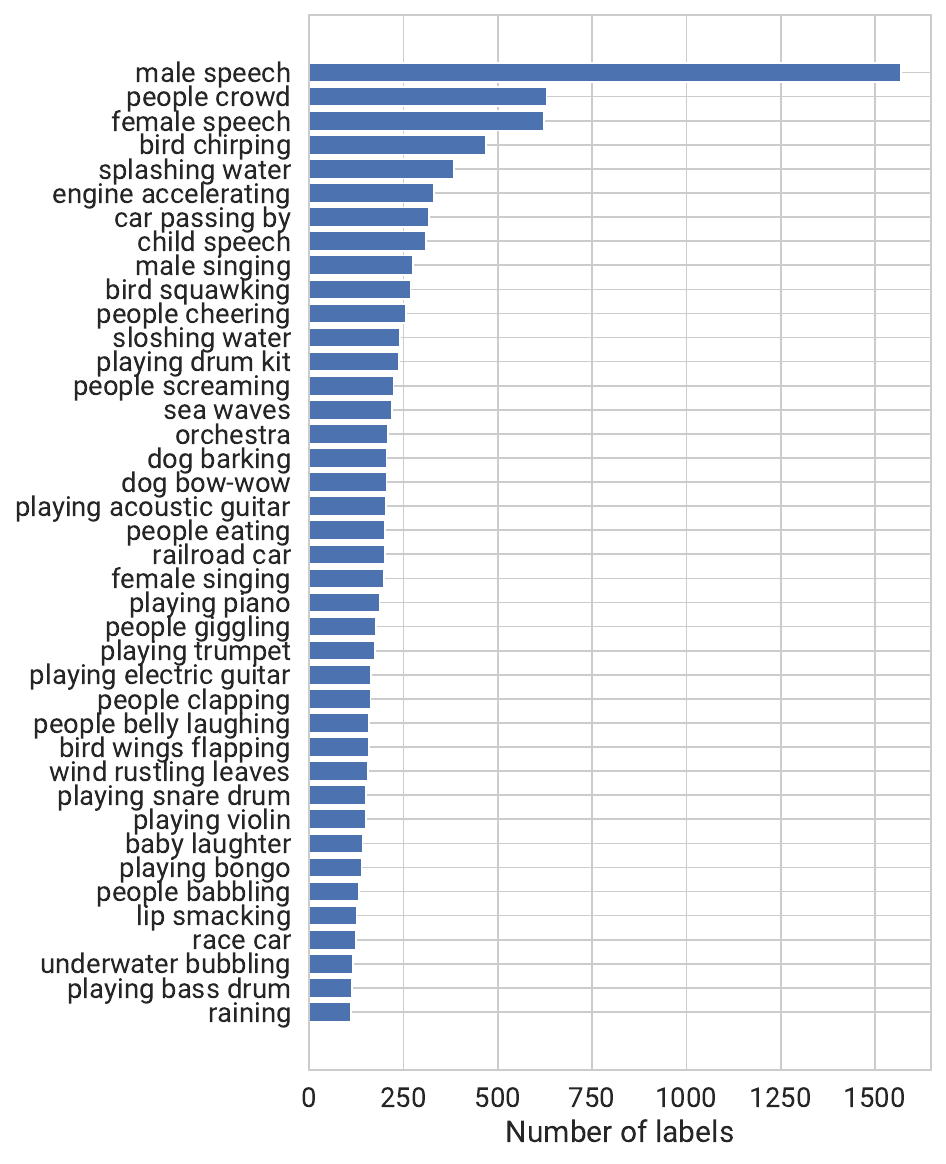}
        \caption{\textit{Visible} classes.}
        \label{fig:visual_freq}
    \end{subfigure}
    \hfill
    \begin{subfigure}[t]{0.32\textwidth}
        \centering
        \includegraphics[width=\linewidth]{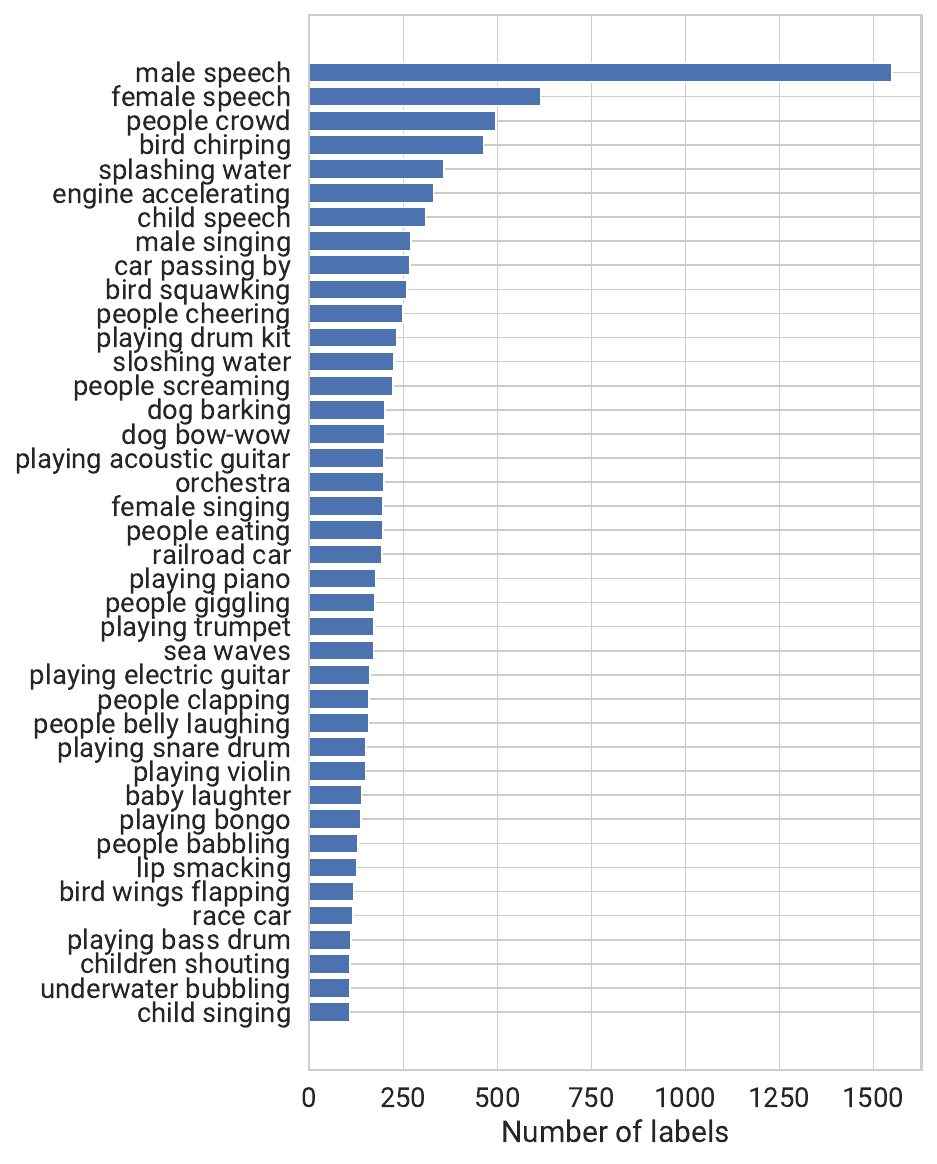}
        \caption{\textit{Audible and visible} classes.}
        \label{fig:audiovisual_freq}
    \end{subfigure}
    \caption{Class label frequency in \vggsounder{} by modality.}
    \label{fig:modality_freqs}
\end{figure}

\noindent \cref{fig:modality_freqs} shows the frequency of the 40 most common class labels by modality. We observe that the label distribution appears to be very similar for visible classes and for classes that are audible and visible. This matches the label modality distribution in Figure 3B in the main paper. Furthermore, we observe that the class label \classname{male speech} occurs more frequently than \classname{female speech}.

\section{Model evaluations and input prompts}
\label{sec:input-prompts}
This section provides additional details about the evaluated models, input prompts and evaluation methodology used in the zero-shot and LLM-assisted evaluations described in \cref{sec:experiments} of the paper. Specifically, we detail the prompts and methods for generating classification predictions for the models in the Gemini family~\cite{team2023gemini} and for the open-source foundational models VideoLLaMA-2~\cite{cheng2024videollama}, Unified-IO 2~\citep{luUnifiedIO2Scaling2023c}, PandaGPT~\citep{suPandaGPTOneModel2023}, and Ola~\citep{liuOlaPushingFrontiers2025}.

\subsection{Models}\label{sec:supp:models}

\noindent \emph{CAV-MAE}~\citep{gong2022contrastive} combines contrastive learning with masked data modelling to obtain strong audio-visual embeddings, used for downstream retrieval and classification tasks. We use the multi-modal CAV-MAE-Scale+ model, pretrained on \audioset{} and fine-tuned on \vggsound{}. Following \cite{gong2022contrastive}, unimodal and multi-modal variants use the original pretrained model but we fine-tune them on \vggsound{} only using the respective modality.\vspace{0.7\baselineskip}

\noindent \emph{DeepAVFusion}~\citep{mo2024unveiling} integrates complementary features from the audio and visual modalities using a deep fusion mechanism, enhancing joint processing for classification tasks. We use publicly available checkpoints for unimodal and multi-modal models pre-trained on \audioset{} and we then fine-tune them on \vggsound{}.\vspace{0.7\baselineskip}

\noindent \emph{AV-Siam}~\citep{lin2024siamese} uses a two-stream network to learn joint embeddings from audio and visual data. By maximising similarity for corresponding pairs and minimising it for non-corresponding pairs, the model captures meaningful relationships between modalities. We use public checkpoints of AV-Siam pre-trained on \audioset{}, to then fine-tune it on \vggsound{}.\vspace{0.7\baselineskip}

\noindent \emph{Equi-AV}~\citep{kim2024equiav} is a transformer-based model that focusses on learning invariant embedding representations through an equivariant learning approach, making it robust to input variations. Again, we fine-tune original model pre-trained on \audioset{} using unimodal or multi-modal \vggsound{} data.\vspace{0.7\baselineskip}

\noindent \emph{Gemini 1.5 Flash, Gemini 1.5 Pro and Gemini 2.0 Flash}~\citep{team2023gemini} are mixture-of-experts transformer models that process both audio and visual information. For classification, the models are prompted to output class labels from the \vggsound{} class list that match the input video clip, along with a caption. Unlike models trained on \vggsound{}, the Gemini models are assumed to be free from \vggsound{}-specific biases. The complete input prompts are provided in \cref{sec:input-prompts}.\vspace{0.7\baselineskip}

\noindent \emph{VideoLLaMA-2.1-AV (VideoLLaMA-2)}~\citep{cheng2024videollama} is a multi-modal foundation model that ingests audio and visual information in two branches that independently process vision-language and audio-language data. The two branches are connected via a language model. VideoLLaMA-2 exhibits strong results on audio-visual question-answering and captioning tasks. Details about the model and prompts used are detailed in \cref{sec:input-prompts}.\vspace{0.7\baselineskip}

\noindent \emph{Unified-IO-2}~\citep{luUnifiedIO2Scaling2023c} is a 7B-parameter autoregressive encoder–decoder model that tokenises text, images, audio, and discrete actions into one shared sequence, enabling ``any-to-any'' understanding and generation.\vspace{0.7\baselineskip}

\noindent \emph{Panda-GPT}~\citep{suPandaGPTOneModel2023} augments a frozen Vicuna-13B language model with ImageBind encoders by using a single linear projection and LoRA adapters. These are trained on only 160k image-text instruction pairs. Despite this lightweight fine-tuning, the model follows instructions across six modalities (image/video, audio, text, depth, thermal, IMU) and can seamlessly compose their semantics in zero-shot settings.\vspace{0.7\baselineskip}

\noindent \emph{Ola}~\citep{liuOlaPushingFrontiers2025} is an omni-modal 7B LLM that progressively aligns modalities—starting with image–text, and then adding speech and finally audio-visual video. It uses local–global attention fusion, dual audio encoders (Whisper~\cite{radford2023robust} + BEATs~\cite{chen2022beats}) and sentence-wise streaming speech decoding. This staged training yields balanced, competitive accuracy for image QA, video QA, and speech recognition.\vspace{0.7\baselineskip}

\mypara{Motivation for LLM-assisted evaluation} \\
In \cref{sec:experiments}, we briefly mentioned standard classification strategies for foundation models, such as: 

\vspace{0.7\baselineskip}
\begin{itemize}
    \item Directly asking for a class without providing a list of available classes \emph{(direct)},
\end{itemize}

\noindent Some models, such as VideoLLaMA-2, Unified-IO-2, and PandaGPT, were pretrained on \vggsound{}. For certain prompts, they return valid \vggsound{} classes, which makes character-level comparison feasible. However, their overall performance on \vggsounder{} is low, as most outputs are synonym classes not included in the original class set.

\vspace{0.7\baselineskip}
\begin{itemize}
    \item Prepending a list of all available classes to the classification prompt \emph{(zero-shot)},
\end{itemize}

\noindent Here, we try to mitigate character-level comparison issues by prepending all 309 class names before the prompt:
\emph{“Annotate the video, explain in detail what is happening in the video. Use classes from the provided list in the captioning and also add yours.”}
This approach works well for closed-source foundation models but performs extremely poorly on all open-source models, most likely due to their smaller effective context window.

\vspace{0.7\baselineskip}
\begin{itemize}
    \item Asking 309 independent questions, one per class, for every sample \emph{(multi-prompt)}.
\end{itemize}

\noindent This strategy avoids the context length limitation. Instead of including all class names at once, we ask 309 questions per sample, each with the prompt:
\emph{“Do you see or hear the following class ‘{class}’ in the video? Answer only with yes or no.”}
While this pipeline yields higher classification scores, it is computationally expensive and still fails to fully capture the video understanding capabilities of most open-source foundation models.

\vspace{0.7\baselineskip}
\noindent In conclusion, all the above strategies yield low performance (e.g., low F1 scores) and fail to reliably capture a model’s video understanding. To address this, we adopt a hybrid approach: we use the \emph{zero-shot} strategy for closed-source models and introduce an \emph{LLM-assisted evaluation} protocol for open-source foundation models.

\paragraph{Gemini models}
The Gemini models can handle long prompts very well. Thus, to generate classification predictions with models from the Gemini family, we used a zero-shot evaluation protocol. Specifically, we provided the models with an input prompt, a list of all class names in \vggsound{} separated with commas, and an input video file. We used the following text template:
\calloutt{\noindent \{CLASSES\} \\ \{VIDEO\} \\ \small\texttt{Annotate the video, explain in detail what is happening in the video. Use classes from the provided list in the captioning and also add yours.}
}

\paragraph{LLM-assisted evaluation} We evaluated all other foundation models using LLM‑assisted evaluation. 

\noindent Building on similar approaches, \footnote{We found the VideoLLaMA-2 appendix \cite{cheng2024videollama}, PointLLM appendix \citep{xuPointLLMEmpoweringLarge2024}, and the Unified-IO 2 code base \url{https://github.com/allenai/unified-io-2/blob/502ac4d81239f82c891a9f412b000c3c8d4e2946/t5x/examples/unified_io/data/prompt_dict.py} to be very useful.
} we employ Qwen3 \citep{yangQwen3TechnicalReport2025} (32B quantised to 8 bits) as our LLM for evaluating the alignment between model-generated outputs and the ground truth. 

\noindent Specifically, for each sample, the open-source foundation models are asked the following questions depending on the input modality:

\calloutt{
\small 
\noindent A: \\ \texttt{
What actions are being performed in this audio, explain all sounds and actions in the audio? Please provide a short answer.
}
}

\calloutt{
\small 
\noindent V/AV: \\ \texttt{
What actions are being performed in this video, explain all sounds and actions in the video? Please provide a short answer.
}
}

\noindent The generated answer (video/audio captioning text) and the target labels (list of classes separated with comma) are then both supplied to the Qwen3 evaluator that receives the following system prompt.

\noindent \mypara{LLM system prompt}

\noindent
\begin{lstlisting}
You are an intelligent chatbot designed for evaluating the correctness of generative outputs for classification pairs. Your task is to compare the predicted answer with the correct answer and determine if they match meaningfully. Here's how you can accomplish the task: 

- Focus on the meaningful match between the predicted answer and the correct answer.
- Consider synonyms or paraphrases as valid matches.
- Evaluate the correctness of the prediction compared to the answer.
- The correct answer, might contain multiple classes. Treat them independently and evaluate the correctness of all them w.r.t predicted answer.

Provide your evaluation only as a yes/no and score where the score is an integer value between 0 and 5, with 5 indicating the highest meaningful match.

Please generate the response in the form of a Python dictionary string where names of classes are keys and values are dictionary strings  with keys 'pred' and 'score', where value of 'pred' is a string of 'yes' or 'no' and value of 'score' is in INTEGER, not STRING.

DO NOT PROVIDE ANY OTHER OUTPUT TEXT OR EXPLANATION. Only provide the Python dictionary string. For example, your response should look like this: 

{"male speech, man speaking": {"pred": "yes", "score": 4}, "playing banjo": {"pred": "no", "score": 0}}

Example 1.

<Question> 
Identify the main sounds present in the given audio clip with a few words.

<Correct Answers> 
["cat caterwauling", "cat meowing"] 

<Predicted Answer> 
The main sounds present in the given audio clip are:

1. A ticking sound, possibly from a clock or timer.
2. A mechanical sound, which could be from a machine or device.
3. A human voice, which is speaking in the background.

Output: {"cat caterwauling": {"pred": "no", "score": 0}, "cat meowing": {"pred": "no", "score": 0}}

Example 2.

<Question> 
What actions are being performed in this audio, explain all sounds and actions in the audio? Please provide a short answer.

<Correct Answers> 
["cuckoo bird calling", "mynah bird singing", "bird chirping, tweeting"]

<Predicted Answer> 
The audio features a cuckoo bird calling in the distance and some chirping and tweeting from smaller birds.

Output: {"cuckoo bird calling": {"pred": "yes", "score": 5}, "mynah bird singing": {"pred": "no", "score": 0}, "bird chirping, tweeting": {"pred": "no", "score": 5}}

Example 3.

<Question> 
What actions are being performed in this video, explain all sounds and actions in the video? Please provide a short answer.

<Correct Answers> 
["male speech, man speaking", "playing hammond organ"]

<Predicted Answer> 
The video shows a man who is playing regular piano and speaking with someone.

Output: {"male speech, man speaking": {"pred": "yes", "score": 5}, "playing hammond organ": {"pred": "yes", "score": 3}}
\end{lstlisting}

\vspace{0.7\baselineskip}

\noindent \mypara{User message template}

\begin{lstlisting}
<Question> 
{QUESTION}

<Correct Answers> 
{ANSWERS}

<Predicted Answer> 
{CAPTION}
\end{lstlisting}

\noindent Qwen3 then outputs a Python‑formatted dictionary mapping for each target class. The dictionary contains binary “pred” decision (yes/no) and a nuanced confidence score (0–5), accommodating synonymy and paraphrasing. 

\calloutt{\small
\noindent\texttt{
\{"male speech, man speaking": \{"pred": "yes", "score": 5\}, "playing hammond organ": \{"pred": "yes", "score": 3\}\}
}
}

\noindent This flexible scoring relaxes the strict label matching, yielding richer, semantically-aware assessments that better reflect human judgment and are aligned with recent ``LLM‑as‑judge''~\citep{gu2025surveyllmasajudge} paradigms that have demonstrated enhanced correlation with human evaluators across a diverse set of tasks and domains.

\begin{figure}[t]
\centering
\begin{minipage}[t]{0.48\textwidth}
    \vspace{0pt} 
     \centering
     \resizebox{\linewidth}{!}{%
        \begin{tabular}{l|ccc|ccc}
        \hline                                                       
        & \multicolumn{3}{c|}{\textbf{Accuracy $\uparrow$}}                                                                & \multicolumn{3}{c}{$\mu \downarrow$ }    \\
        \cmidrule(lr){2-4} \cmidrule(lr){5-7} 
        \bf{Models} & \multicolumn{1}{c}{\textit{a}}  & \multicolumn{1}{c}{\textit{v}}   & \multicolumn{1}{c|}{\textit{av}}  & \multicolumn{1}{c}{\textit{$\mu_A$}}  & \multicolumn{1}{c}{\textit{$\mu_V$}}   & \multicolumn{1}{c}{$\mu_{A \cap V}$}              \\
        \hline
CAV-MAE & 59.05 & 45.57 & 65.08 & 4.71 & 4.84 & 0.67 \\
DeepAVFusion & 40.82 & 27.24 & 53.10 & 4.18 & 3.17 & 0.07 \\
Equi-AV & 46.68 & 24.84 & 50.08 & 6.91 & 5.51 & 0.98 \\
AV-Siam & 56.91 & 47.27 & 55.25 & 13.17 & 8.92 & 3.92 \\
        \midrule
Gemini 1.5 Flash & 0.31 & 22.12 & 23.60 & 1.51 & 4.17 & 0.09 \\
Gemini 1.5 Pro & 1.29 & 25.77 & 21.31 & 1.62 & 5.41 & 0.24 \\
Gemini 2.0 Flash & 5.70 & 20.29 & 19.39 & 2.50 & 4.77 & 0.63 \\
\midrule
VideoLLaMA 2 & 27.98 & 17.01 & 21.46 & 11.16 & 2.85 & 1.42 \\
Unified-IO 2 & 32.28 & 20.24 & 52.40 & 4.88 & 3.42 & 0.87 \\
PandaGPT & 5.20 & 7.65 & 8.95 & 4.51 & 4.48 & 0.94 \\
OLA & 10.71 & 8.63 & 14.29 & 7.61 & 4.05 & 0.71 \\
        \hline
        \end{tabular}
    }
    \caption{\textbf{Performance of state-of-the-art models on \vggsound{}.} We report top-1 classification accuracy for different input modalities (audio \textit{A}, visual \textit{V}, and audio and visual information \textit{AV}). $\mu$ is modality confusion metric defined in \cref{sec:experiments}}
    \label{tab:vggsound_test_table}
\end{minipage}
\hfill
\begin{minipage}[t]{0.48\textwidth}
    \vspace{0pt}
    \centering
    \includegraphics[width=\linewidth]{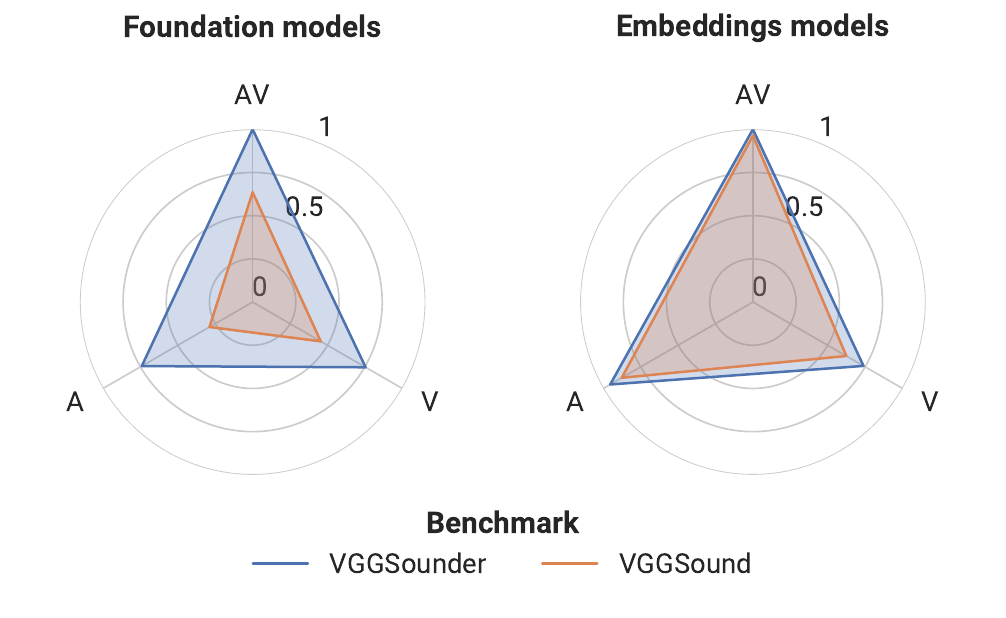}
    \captionof{figure}{\textbf{Performance of state-of-the-art families on \vggsound{} compared to \vggsounder{}.} Radar plots illustrate the average F1-scores across modalities for two model families: ``Foundation models'' and ``Embedding models'' (\cref{tab:main}).}
    \label{fig:families-radar-f1}
\end{minipage}

\end{figure}

\section{Additional quantitative analysis}
\label{sec:additional-quantitative-analysis}

This appendix extends our quantitative analyses presented in \cref{sec:reeval-sota} of the main paper, providing further insights into model behaviour on both \vggsound{} and the newly introduced \vggsounder{} benchmark.

\newpage

\subsection{Model performance on \vggsound{}}

\noindent
\begin{wrapfigure}[27]{r}{0.55\textwidth} 
    \centering
    \vspace{-45pt} 
    \includegraphics[width=0.62\linewidth]{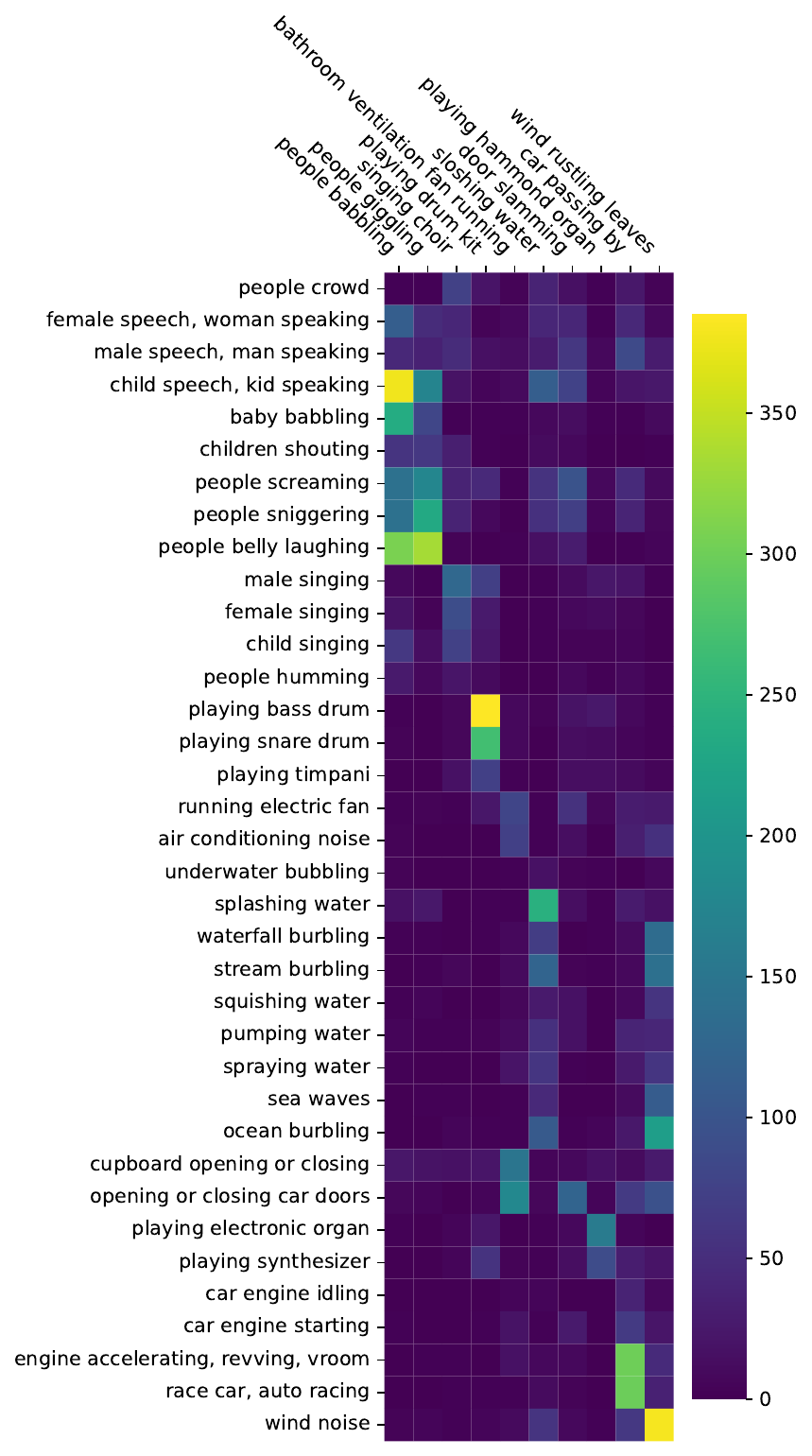}
    \vspace{-10pt} 
    \caption{Co-occurrence counts among a subset of VGGSound classes, estimated on the VGGSound test set by the CAV-MAE model. Each cell indicates how frequently two classes appear together, highlighting labels that share overlapping acoustic cues (e.g., \classname{playing drum kit} and \classname{playing bass drum}). Best viewed zoomed in on a screen.}
    \label{fig:confusion_matrix}
\end{wrapfigure}

\noindent We present the classification performance of state-of-the-art models on the original \vggsound{} test data in \cref{tab:vggsound_test_table}. We observe that the multi-label hit accuracy on \vggsounder{} reported in \cref{tab:main} in the main paper significantly raises the performance across all models. This suggests that the models predict classes that were not present in the original \vggsound{} labelling, despite those being correct. 

\noindent \cref{fig:families-radar-f1} further compares the averaged F1-scores between the ``Foundation model'' and ``Embedding model'' families, highlighting that evaluations on the original \vggsound{} consistently underestimate model performance across all modalities when compared to evaluations on \vggsounder{}.

\subsection{Co-occurrence matrix on \vggsound{}}

To further illustrate the issue of class overlap described in \cref{sec:vggsound-issues} of our paper, we include an analysis of class co-occurrences in predictions by the CAV-MAE ~\citep{gong2022contrastive} model in \cref{fig:confusion_matrix}. Specifically, we provide a co-occurrence matrix highlighting frequent simultaneous predictions of certain classes. Notably, labels such as \classname{playing drum kit} and \classname{playing bass drum} are frequently predicted together, as they are not mutually exclusive. This analysis supports our identification of overlapping classes as a key limitation in the original \vggsound{} annotations and demonstrates the need for explicitly multi-label approaches in video classification tasks.

\subsection{Classification results for other $k$}
\cref{tab:av_vggsound_allk_vertical} extends the evaluation presented in the main paper by showing multi-label video classification results on \vggsounder{} for varying numbers of top-$k$ predictions, specifically for $k \in \{3, 5, 10\}$. These additional results offer deeper insights into how model performance changes with an increasing number of predictions. Specifically, one can notice the opposite behaviour between the F1-score (goes down with $k$) and the Hit score (increases with $k$). 

\begin{table*}[th]
  \centering

  \resizebox{0.7\textwidth}{!}{
  \begin{tabular}{
    l                                   
    l                                   
    S[table-format=1.2]S[table-format=1.2]S[table-format=1.2]  
    S[table-format=2.2]S[table-format=2.2]S[table-format=2.2]  
    S[table-format=2.2]S[table-format=2.2]S[table-format=2.2]} 
  \toprule
  & & \multicolumn{3}{c}{\textbf{Subset Accuracy}\,$\uparrow$} &
      \multicolumn{3}{c}{$\mathbf{F_1}$\,$\uparrow$} &
      \multicolumn{3}{c}{\textbf{Hit}\,$\uparrow$} \\
  \cmidrule(lr){3-5}\cmidrule(lr){6-8}\cmidrule(lr){9-11}
  \textbf{$k$} & \textbf{Model} &
  \emph{a} & \emph{v} & \emph{av} &
  \emph{a} & \emph{v} & \emph{av} &
  \emph{a} & \emph{v} & \emph{AV} \\
  \midrule
  \multirow{4}{*}{3}
& CAV-MAE & 1.23 & 0.71 & 1.41 & 41.96 & 37.93 & 45.35 & 83.18 & 74.42 & 84.15 \\
& DeepAVFusion & 0.31 & 0.05 & 0.85 & 30.49 & 23.85 & 39.78 & 67.88 & 52.68 & 77.09 \\
& Equi-AV & 0.74 & 0.24 & 0.59 & 36.25 & 24.11 & 36.57 & 76.34 & 49.66 & 73.45 \\
& AV-Siam & 1.17 & 0.87 & 1.04 & 40.15 & 39.31 & 43.07 & 81.30 & 75.50 & 81.67 \\
  \midrule
  \multirow{4}{*}{5}
& CAV-MAE & 0.03 & 0.01 & 0.01 & 36.32 & 32.09 & 37.01 & 87.95 & 80.48 & 88.56 \\
& DeepAVFusion & 0.00 & 0.00 & 0.01 & 26.45 & 20.39 & 32.46 & 74.58 & 60.49 & 82.42 \\
& Equi-AV & 0.01 & 0.00 & 0.00 & 31.72 & 21.18 & 30.01 & 82.66 & 57.13 & 79.13 \\
& AV-Siam & 0.01 & 0.02 & 0.02 & 34.62 & 32.99 & 35.54 & 86.06 & 81.20 & 86.71 \\
  \midrule
  \multirow{4}{*}{10}
& CAV-MAE & 0.00 & 0.00 & 0.00 & 25.67 & 22.00 & 24.35 & 92.16 & 86.69 & 92.74 \\
& DeepAVFusion & 0.00 & 0.00 & 0.00 & 18.99 & 14.58 & 21.48 & 82.16 & 69.64 & 87.41 \\
& Equi-AV & 0.00 & 0.00 & 0.00 & 22.91 & 15.43 & 20.25 & 88.73 & 66.67 & 85.33 \\
& AV-Siam & 0.00 & 0.00 & 0.00 & 24.42 & 22.46 & 23.90 & 90.70 & 87.39 & 92.01 \\
  \bottomrule
  \end{tabular}
  }
  \caption{\textbf{Audio-visual video classification results on \vggsounder{} for $k\!\in\!\{3,5,10\}$.}
           The table is vertically grouped by $k$. Within each block, the four models are compared
           across the three metrics and input modalities.}
  \label{tab:av_vggsound_allk_vertical}
\end{table*}

\subsection{Performance on subsets of \vggsounder{}}\label{sec:confounds-performance}
To comprehensively evaluate model robustness in the presence of common confounders highlighted in \cref{sec:vggsound-issues} (i.e.\ meta-labels: \textit{background music}, \textit{static images}, and \textit{voice over}), we present additional evaluations on distinct subsets of \vggsounder{}. Specifically, \cref{tab:bg_music,tab:no_bg,tab:static_images,tab:no_static_images,tab:voice_over,tab:no_voice_over} display the performance of state-of-the-art models on subsets only containing or fully excluding the meta-labels. These analyses confirm the importance of accounting for modality-specific and meta-label influences.

\mypara{Impact of background music}

\begin{table*}[ht!]
\centering
\resizebox{\textwidth}{!}{
\begin{tabular}{l S[table-format=2.2] S[table-format=2.2] S[table-format=2.2] S[table-format=2.2] S[table-format=2.2] S[table-format=2.2] S[table-format=2.2] S[table-format=2.2] S[table-format=2.2] S[table-format=2.2] S[table-format=2.2] S[table-format=2.2] S[table-format=2.2] S[table-format=2.2]}
\toprule
& \multicolumn{3}{c}{\textbf{Subset Accuracy} \small$\uparrow$} & \multicolumn{5}{c}{$\mathbf{F_1}$ \small$\uparrow$} & \multicolumn{3}{c}{\textbf{Hit} \small$\uparrow$} & \multicolumn{3}{c}{$\mathbf{\mu}$ \small$\downarrow$} \\
\cmidrule(lr){2-4} \cmidrule(lr){5-9} \cmidrule(lr){10-12} \cmidrule(lr){13-15}
\textbf{Model} & \emph{a} & \emph{v} & \emph{av} & \emph{a} & \emph{v} & \emph{av} &   \emph{$a(A\neg V)$} &   \emph{$v(V\neg A)$} & \emph{a} & \emph{v} & \emph{av} & \emph{$\mu_A$} & \emph{$\mu_V$} & \emph{$\mu_{A \cap V}$} \\
\midrule
CAV-MAE & 11.94 & 24.31 & 28.21 & 32.73 & 36.60 & 43.78 & 20.34 & 31.28 & 57.40 & 49.66 & 58.61 & 5.27 & 4.58 & 0.59 \\
DeepAVFusion & 8.35 & 12.33 & 24.07 & 21.68 & 19.64 & 36.37 & 13.98 & 18.14 & 38.05 & 26.65 & 48.68 & 3.30 & 2.82 & 0.10 \\
Equi-AV & 9.84 & 12.86 & 22.29 & 26.07 & 20.08 & 34.22 & 16.48 & 15.23 & 45.72 & 27.24 & 45.80 & 7.79 & 5.11 & 1.01 \\
AV-Siam & 11.15 & 23.09 & 24.09 & 30.96 & 35.17 & 37.73 & 19.17 & 35.80 & 54.31 & 47.72 & 50.50 & 13.88 & 7.04 & 3.14 \\

\midrule

Gemini 1.5 Flash & 2.57 & 18.04 & 17.35 & 15.58 & 38.57 & 42.47 & 13.12 & 33.89 & 34.51 & 47.14 & 56.66 & 17.25 & 3.34 & 1.18 \\
Gemini 1.5 Pro & 4.08 & 28.01 & 27.12 & 20.88 & 54.04 & 56.38 & 16.47 & 37.92 & 38.75 & 71.67 & 77.57 & 5.63 & 3.67 & 0.95 \\
Gemini 2.0 Flash & 1.22 & 14.70 & 12.04 & 12.08 & 35.97 & 36.31 & 10.65 & 32.18 & 19.24 & 41.63 & 45.85 & 3.40 & 3.70 & 0.98 \\

\midrule

VideoLLaMA 2 & 12.37 & 23.32 & 26.98 & 36.60 & 47.94 & 51.63 & 24.60 & 44.67 & 51.81 & 47.72 & 52.49 & 20.62 & 4.78 & 2.78 \\
Unified-IO 2 & 9.24 & 16.64 & 27.64 & 27.81 & 33.63 & 47.68 & 20.24 & 28.92 & 39.34 & 31.30 & 53.44 & 9.62 & 4.68 & 1.41 \\
PandaGPT & 2.53 & 6.45 & 7.63 & 14.46 & 20.56 & 21.24 & 9.89 & 19.57 & 16.74 & 16.91 & 16.74 & 9.66 & 6.12 & 2.72 \\
OLA & 11.22 & 12.13 & 22.52 & 39.89 & 29.62 & 48.64 & 31.72 & 27.17 & 48.52 & 26.79 & 50.02 & 14.17 & 6.35 & 2.19 \\

\bottomrule
\end{tabular}
}
\caption{%
\textbf{Audio-visual video classification results on the subset of \vggsounder{} that is labelled as containing \textit{background music}.} 
Similar to Table 1 in the main paper, we report multi-label classification metrics (subset accuracy, $F_1$-score, Hit accuracy, modality confusion ($\mu$)) for audio- $a(A)$, visual - $v(V)$, audio-visual - $av(AV)$, audio-only - $a(A\neg V)$ and video-only - $v(V\neg A)$ inputs.}
\label{tab:bg_music}
\end{table*}

\begin{table*}[ht!]
\centering
\resizebox{\textwidth}{!}{
\begin{tabular}{l S[table-format=2.2] S[table-format=2.2] S[table-format=2.2] S[table-format=2.2] S[table-format=2.2] S[table-format=2.2] S[table-format=2.2] S[table-format=2.2] S[table-format=2.2] S[table-format=2.2] S[table-format=2.2] S[table-format=2.2] S[table-format=2.2] S[table-format=2.2]}
\toprule
& \multicolumn{3}{c}{\textbf{Subset Accuracy} \small$\uparrow$} & \multicolumn{5}{c}{$\mathbf{F_1}$ \small$\uparrow$} & \multicolumn{3}{c}{\textbf{Hit} \small$\uparrow$} & \multicolumn{3}{c}{$\mathbf{\mu}$ \small$\downarrow$} \\
\cmidrule(lr){2-4} \cmidrule(lr){5-9} \cmidrule(lr){10-12} \cmidrule(lr){13-15}
\textbf{Model} & \emph{a} & \emph{v} & \emph{av} & \emph{a} & \emph{v} & \emph{av} &   \emph{$a(A\neg V)$} &   \emph{$v(V\neg A)$} & \emph{a} & \emph{v} & \emph{av} & \emph{$\mu_A$} & \emph{$\mu_V$} & \emph{$\mu_{A \cap V}$} \\
\midrule

CAV-MAE & 17.94 & 24.21 & 29.61 & 39.53 & 38.87 & 47.24 & 21.69 & 23.56 & 67.07 & 56.37 & 67.51 & 4.07 & 4.93 & 0.85 \\
DeepAVFusion & 13.89 & 13.83 & 26.16 & 29.06 & 23.29 & 41.81 & 15.67 & 13.65 & 49.40 & 33.80 & 59.79 & 4.21 & 3.42 & 0.26 \\
Equi-AV & 15.80 & 13.57 & 24.40 & 33.65 & 22.93 & 39.33 & 19.02 & 14.39 & 57.09 & 33.25 & 56.21 & 7.77 & 6.28 & 1.51 \\
AV-Siam & 17.40 & 24.99 & 27.86 & 38.12 & 39.77 & 43.73 & 20.41 & 23.77 & 64.69 & 57.68 & 62.49 & 10.76 & 8.55 & 3.98 \\

\midrule

Gemini 1.5 Flash & 2.36 & 18.55 & 20.24 & 14.21 & 42.02 & 46.82 & 17.06 & 30.96 & 30.75 & 51.79 & 62.30 & 13.25 & 4.12 & 1.45 \\
Gemini 1.5 Pro & 3.48 & 27.18 & 26.91 & 19.38 & 54.90 & 57.20 & 20.02 & 32.14 & 33.43 & 73.27 & 78.36 & 2.65 & 4.03 & 0.52 \\
Gemini 2.0 Flash & 2.43 & 16.31 & 15.48 & 12.86 & 38.67 & 40.50 & 7.73 & 26.45 & 18.86 & 47.01 & 50.56 & 2.59 & 4.79 & 1.02 \\

\midrule

VideoLLaMA 2 & 16.25 & 23.88 & 27.70 & 40.76 & 50.28 & 53.67 & 25.55 & 38.01 & 58.15 & 53.22 & 59.92 & 14.51 & 5.43 & 3.11 \\
Unified-IO 2 & 15.27 & 14.65 & 29.71 & 37.69 & 30.51 & 50.10 & 27.43 & 21.30 & 54.43 & 32.18 & 63.36 & 10.80 & 4.64 & 1.84 \\
PandaGPT & 3.40 & 5.35 & 6.30 & 18.58 & 19.39 & 21.55 & 17.66 & 18.73 & 19.72 & 17.10 & 18.74 & 9.45 & 6.33 & 3.03 \\
OLA & 15.99 & 10.54 & 19.00 & 48.47 & 26.52 & 46.04 & 45.53 & 17.35 & 57.50 & 24.84 & 50.20 & 17.34 & 6.23 & 2.85 \\

\bottomrule
\end{tabular}
}
\caption{\textbf{Audio-visual video classification results on the subset of \vggsounder{} that is labelled as \textit{not} containing \textit{background music}}}\label{tab:no_bg}
\end{table*}

\noindent A side-by-side inspection of the two subsets (Tab.\ref{tab:bg_music} vs.\ Tab.\ref{tab:no_bg}) reveals several interesting points. 

\vspace{0.7\baselineskip}
\noindent \textit{(i) Universal but modality-specific gains.}  Every method improves in terms of $F_1$ and \emph{Hit} scores when the soundtrack is removed, that is especially clear for the \textit{audio} input modality: for the embedding family we register jumps of up to $+5\%$ in $F_1$ for both audio and \textit{visual} inputs.  Consequently, joint audio–visual inputs rise in performance only slightly ($+3$–$5\%$).

\vspace{0.7\baselineskip}
\noindent \textit{(ii) Same trend for foundation models, but with caveats.}  Foundation checkpoints with a meaningful audio encoder echo the pattern (Unified-IO2 $+7\%$, Ola $+12\%$); in contrast, the Gemini family remains audio-weak, suggesting that their publicly released models rely heavily on vision.  

\vspace{0.7\baselineskip}
\noindent \textit{(iii) Intuition.}  Background music tends to mask class-specific foreground sounds; once that mask is removed the audio encoder can finally ``hear'' discriminative cues, whereas vision—being agnostic to the soundtrack—is affected only by the changed clip mix.  With noisy audio, every model relies more on the \textit{V} modality as a safety net, which explains why their baseline performance remains respectable despite the severe audio corruption.  

\vspace{0.7\baselineskip}
\noindent Altogether, these observations confirm that background music constitutes a hard confounder, forcing models to rely on vision.

\vspace{0.7\baselineskip}
\mypara{Impact of static images}

\begin{table*}[ht!]
\centering
\resizebox{\textwidth}{!}{
\begin{tabular}{l S[table-format=2.2] S[table-format=2.2] S[table-format=2.2] S[table-format=2.2] S[table-format=2.2] S[table-format=2.2] S[table-format=2.2] S[table-format=2.2] S[table-format=2.2] S[table-format=2.2] S[table-format=2.2] S[table-format=2.2] S[table-format=2.2] S[table-format=2.2]}
\toprule
& \multicolumn{3}{c}{\textbf{Subset Accuracy} \small$\uparrow$} & \multicolumn{5}{c}{$\mathbf{F_1}$ \small$\uparrow$} & \multicolumn{3}{c}{\textbf{Hit} \small$\uparrow$} & \multicolumn{3}{c}{$\mathbf{\mu}$ \small$\downarrow$} \\
\cmidrule(lr){2-4} \cmidrule(lr){5-9} \cmidrule(lr){10-12} \cmidrule(lr){13-15}
\textbf{Model} & \emph{a} & \emph{v} & \emph{av} & \emph{a} & \emph{v} & \emph{av} &   \emph{$a(A\neg V)$} &   \emph{$v(V\neg A)$} & \emph{a} & \emph{v} & \emph{av} & \emph{$\mu_A$} & \emph{$\mu_V$} & \emph{$\mu_{A \cap V}$} \\
\midrule
CAV-MAE & 25.31 & 26.75 & 33.76 & 42.56 & 33.68 & 42.08 & 39.94 & 24.11 & 66.50 & 41.64 & 51.29 & 5.67 & 3.27 & 0.49 \\
DeepAVFusion & 18.43 & 15.40 & 28.26 & 30.47 & 17.47 & 35.46 & 28.76 & 10.61 & 47.68 & 21.59 & 43.14 & 4.48 & 2.75 & 0.26 \\
Equi-AV & 22.46 & 11.25 & 26.21 & 36.68 & 15.37 & 33.91 & 34.22 & 8.51 & 57.32 & 19.00 & 41.32 & 9.00 & 3.58 & 0.68 \\
AV-Siam & 24.50 & 23.71 & 30.06 & 40.73 & 29.75 & 38.92 & 37.55 & 22.70 & 63.65 & 36.78 & 47.43 & 16.65 & 5.43 & 2.84 \\

\midrule

Gemini 1.5 Flash & 2.11 & 19.60 & 22.99 & 10.47 & 39.35 & 43.59 & 9.79 & 28.97 & 22.02 & 43.01 & 50.96 & 12.52 & 3.82 & 0.74 \\
Gemini 1.5 Pro & 3.35 & 32.83 & 32.64 & 15.64 & 54.04 & 56.49 & 14.49 & 36.26 & 25.43 & 64.74 & 70.90 & 4.99 & 4.01 & 0.86 \\
Gemini 2.0 Flash & 4.34 & 18.69 & 17.20 & 13.91 & 35.67 & 38.01 & 13.34 & 23.19 & 19.29 & 38.91 & 43.41 & 3.27 & 3.21 & 0.92 \\

\midrule

VideoLLaMA 2 & 21.77 & 25.99 & 33.60 & 44.07 & 47.01 & 52.88 & 40.51 & 35.66 & 57.26 & 42.10 & 51.13 & 23.12 & 3.51 & 2.47 \\
Unified-IO 2 & 20.29 & 14.89 & 33.60 & 37.18 & 27.76 & 48.92 & 35.52 & 17.58 & 47.21 & 22.95 & 50.16 & 9.06 & 3.08 & 1.23 \\
PandaGPT & 2.30 & 6.99 & 8.36 & 11.98 & 16.68 & 20.31 & 10.92 & 11.24 & 12.28 & 12.31 & 15.27 & 7.71 & 4.56 & 1.73 \\
OLA & 16.87 & 8.66 & 22.83 & 40.52 & 19.98 & 43.52 & 36.85 & 14.74 & 44.73 & 15.05 & 42.12 & 17.45 & 3.39 & 1.17 \\

\bottomrule
\end{tabular}
}
\caption{\textbf{Audio-visual video classification results on the subset of \vggsounder{} that is labelled as containing \textit{static images}}}\label{tab:static_images}
\end{table*}

\begin{table*}[ht!]
\centering
\resizebox{\textwidth}{!}{
\begin{tabular}{l S[table-format=2.2] S[table-format=2.2] S[table-format=2.2] S[table-format=2.2] S[table-format=2.2] S[table-format=2.2] S[table-format=2.2] S[table-format=2.2] S[table-format=2.2] S[table-format=2.2] S[table-format=2.2] S[table-format=2.2] S[table-format=2.2] S[table-format=2.2]}
\toprule
& \multicolumn{3}{c}{\textbf{Subset Accuracy} \small$\uparrow$} & \multicolumn{5}{c}{$\mathbf{F_1}$ \small$\uparrow$} & \multicolumn{3}{c}{\textbf{Hit} \small$\uparrow$} & \multicolumn{3}{c}{$\mathbf{\mu}$ \small$\downarrow$} \\
\cmidrule(lr){2-4} \cmidrule(lr){5-9} \cmidrule(lr){10-12} \cmidrule(lr){13-15}
\textbf{Model} & \emph{a} & \emph{v} & \emph{av} & \emph{a} & \emph{v} & \emph{av} &   \emph{$a(A\neg V)$} &   \emph{$v(V\neg A)$} & \emph{a} & \emph{v} & \emph{av} & \emph{$\mu_A$} & \emph{$\mu_V$} & \emph{$\mu_{A \cap V}$} \\
\midrule

CAV-MAE & 15.68 & 24.09 & 29.12 & 37.62 & 38.72 & 46.88 & 17.53 & 25.24 & 64.94 & 55.93 & 66.75 & 4.15 & 5.05 & 0.83 \\
DeepAVFusion & 12.07 & 13.46 & 25.65 & 27.18 & 22.91 & 41.15 & 12.43 & 14.86 & 46.99 & 33.12 & 58.62 & 3.97 & 3.37 & 0.22 \\
Equi-AV & 13.64 & 13.57 & 23.91 & 31.56 & 22.79 & 38.70 & 15.09 & 14.96 & 54.48 & 32.92 & 55.10 & 7.63 & 6.34 & 1.49 \\
AV-Siam & 15.12 & 24.70 & 27.04 & 36.18 & 39.44 & 42.91 & 16.51 & 26.53 & 62.45 & 56.97 & 61.09 & 10.76 & 8.58 & 3.93 \\

\midrule

Gemini 1.5 Flash & 2.44 & 18.40 & 19.56 & 14.94 & 41.52 & 46.20 & 17.29 & 31.74 & 32.66 & 51.40 & 61.88 & 14.25 & 3.98 & 1.47 \\
Gemini 1.5 Pro & 3.63 & 27.02 & 26.64 & 20.12 & 54.78 & 57.09 & 19.93 & 33.15 & 35.61 & 73.45 & 78.62 & 3.04 & 3.95 & 0.58 \\
Gemini 2.0 Flash & 1.93 & 15.88 & 14.75 & 12.56 & 38.34 & 39.87 & 7.56 & 27.88 & 18.90 & 46.45 & 50.08 & 2.69 & 4.73 & 1.03 \\

\midrule

VideoLLaMA 2 & 14.69 & 23.66 & 27.24 & 39.46 & 50.03 & 53.36 & 22.14 & 39.43 & 56.82 & 52.80 & 59.02 & 14.86 & 5.51 & 3.12 \\
Unified-IO 2 & 13.29 & 15.02 & 29.11 & 35.55 & 31.17 & 49.75 & 23.50 & 23.14 & 51.86 & 32.53 & 62.25 & 10.74 & 4.83 & 1.81 \\
PandaGPT & 3.33 & 5.47 & 6.43 & 18.31 & 19.70 & 21.55 & 16.45 & 19.35 & 19.94 & 17.33 & 18.56 & 9.70 & 6.50 & 3.12 \\
OLA & 14.80 & 10.95 & 19.44 & 47.37 & 27.33 & 46.59 & 42.96 & 19.62 & 57.00 & 25.76 & 50.61 & 16.61 & 6.60 & 2.90 \\

\bottomrule
\end{tabular}
}
\caption{\textbf{Audio-visual video classification results on the subset of \vggsounder{} that is labelled as \textit{not} containing \textit{static images}}}\label{tab:no_static_images}
\end{table*}

\noindent A side-by-side inspection of the ``static image'' split (Tab.,\ref{tab:static_images}) and its complement (Tab.,\ref{tab:no_static_images}) shows four salient effects.

\vspace{0.7\baselineskip}
\noindent \textit{(i) Vision takes the hit, audio steps up.}  Across the classic embedding models, the \textit{visual} branch loses on average $6$–$7\%$ absolute in $F_1$, while using \textit{audio} inputs results in gains $+3$–$6\%$. The same holds true for the joint audio-visual.  Hit scores mirror the trend: Hit for visual inputs plunges by up to $20\%$, whereas Hit for audio inputs remains flat or edges upward for most of the models.

\vspace{0.7\baselineskip}
\noindent \textit{(ii) Foundation models react unevenly.} VideoLLaMA2 loses $~8\%$ on vision yet gains $~3\%$ on audio—whereas the vision-centric Gemini family suffers a broad decline, unable to compensate for the poor visual signal.

\vspace{0.7\baselineskip}
\noindent \textit{(iii) Intuition.}  Static clips provide far less discriminative visual evidence than genuine video, reducing motion and viewpoint cues.  The audio track, in contrast, is untouched; consequently, models shift their reliance toward the acoustic channel, explaining the systematic audio gain and the parallel vision loss.

\vspace{0.7\baselineskip}
\noindent \textit{(iv) Modality-confusion drifts upward.}  With vision degraded, many architectures become more uncertain about which modality to trust; a few (AV-Siam) even over-correct, raising $\mu_{A}$ by $+1.8$ while slightly easing $\mu_{V}$.

\vspace{0.7\baselineskip}
\noindent In sum, static imagery acts as the visual analogue to background music: it removes discriminative content in one modality (vision) and forces models to lean on the other (audio), exposing how well a model can rebalance modalities.

\vspace{0.7\baselineskip}
\mypara{Impact of voice-over narration}

\begin{table*}[ht!]
\centering
\resizebox{\textwidth}{!}{
\begin{tabular}{l S[table-format=2.2] S[table-format=2.2] S[table-format=2.2] S[table-format=2.2] S[table-format=2.2] S[table-format=2.2] S[table-format=2.2] S[table-format=2.2] S[table-format=2.2] S[table-format=2.2] S[table-format=2.2] S[table-format=2.2] S[table-format=2.2] S[table-format=2.2]}
\toprule
& \multicolumn{3}{c}{\textbf{Subset Accuracy} \small$\uparrow$} & \multicolumn{5}{c}{$\mathbf{F_1}$ \small$\uparrow$} & \multicolumn{3}{c}{\textbf{Hit} \small$\uparrow$} & \multicolumn{3}{c}{$\mathbf{\mu}$ \small$\downarrow$} \\
\cmidrule(lr){2-4} \cmidrule(lr){5-9} \cmidrule(lr){10-12} \cmidrule(lr){13-15}
\textbf{Model} & \emph{a} & \emph{v} & \emph{av} & \emph{a} & \emph{v} & \emph{av} &   \emph{$a(A\neg V)$} &   \emph{$v(V\neg A)$} & \emph{a} & \emph{v} & \emph{av} & \emph{$\mu_A$} & \emph{$\mu_V$} & \emph{$\mu_{A \cap V}$} \\
\midrule
CAV-MAE & 2.59 & 21.69 & 26.64 & 28.79 & 35.34 & 43.14 & 12.64 & 29.17 & 54.06 & 49.85 & 59.77 & 4.76 & 5.16 & 0.74 \\
DeepAVFusion & 1.57 & 13.91 & 21.37 & 18.07 & 23.06 & 34.89 & 8.22 & 18.09 & 34.04 & 32.63 & 48.47 & 3.89 & 3.44 & 0.14 \\
Equi-AV & 3.12 & 11.77 & 19.59 & 23.69 & 20.42 & 32.55 & 12.09 & 18.15 & 44.49 & 28.81 & 45.09 & 9.25 & 6.37 & 1.37 \\
AV-Siam & 2.45 & 22.21 & 23.36 & 26.90 & 35.82 & 37.70 & 11.30 & 30.15 & 50.50 & 50.54 & 52.23 & 11.59 & 8.24 & 3.79 \\

\midrule

Gemini 1.5 Flash & 7.76 & 16.04 & 17.18 & 33.69 & 37.92 & 41.74 & 36.97 & 35.44 & 70.08 & 45.58 & 55.73 & 34.10 & 4.25 & 3.18 \\
Gemini 1.5 Pro & 9.97 & 23.98 & 15.68 & 38.19 & 50.95 & 51.06 & 38.57 & 34.71 & 73.47 & 67.44 & 76.73 & 6.73 & 2.55 & 1.37 \\
Gemini 2.0 Flash & 0.34 & 14.88 & 13.18 & 11.21 & 36.15 & 36.49 & 5.01 & 30.96 & 18.33 & 44.76 & 46.77 & 2.95 & 4.66 & 1.14 \\

\midrule

VideoLLaMA 2 & 5.47 & 22.81 & 26.68 & 33.96 & 49.33 & 52.51 & 21.25 & 42.67 & 50.87 & 51.75 & 57.14 & 16.18 & 5.76 & 3.02 \\
Unified-IO 2 & 6.62 & 14.32 & 26.45 & 33.02 & 30.35 & 46.79 & 28.98 & 24.77 & 53.16 & 30.27 & 56.00 & 17.45 & 4.25 & 1.61 \\
PandaGPT & 4.13 & 5.43 & 6.27 & 23.30 & 20.17 & 21.41 & 24.02 & 18.22 & 29.25 & 17.77 & 17.68 & 15.71 & 8.17 & 4.66 \\
OLA & 20.05 & 9.79 & 13.05 & 58.43 & 25.75 & 45.65 & 57.41 & 24.02 & 80.76 & 24.58 & 55.14 & 17.49 & 5.06 & 3.99 \\

\bottomrule
\end{tabular}
}
\caption{\textbf{Audio-visual video classification results on the subset of \vggsounder{} that is labelled as containing \textit{voice over} narrations}}\label{tab:voice_over}
\end{table*}

\begin{table*}[ht!]
\centering
\resizebox{\textwidth}{!}{
\begin{tabular}{l S[table-format=2.2] S[table-format=2.2] S[table-format=2.2] S[table-format=2.2] S[table-format=2.2] S[table-format=2.2] S[table-format=2.2] S[table-format=2.2] S[table-format=2.2] S[table-format=2.2] S[table-format=2.2] S[table-format=2.2] S[table-format=2.2] S[table-format=2.2]}
\toprule
& \multicolumn{3}{c}{\textbf{Subset Accuracy} \small$\uparrow$} & \multicolumn{5}{c}{$\mathbf{F_1}$ \small$\uparrow$} & \multicolumn{3}{c}{\textbf{Hit} \small$\uparrow$} & \multicolumn{3}{c}{$\mathbf{\mu}$ \small$\downarrow$} \\
\cmidrule(lr){2-4} \cmidrule(lr){5-9} \cmidrule(lr){10-12} \cmidrule(lr){13-15}
\textbf{Model} & \emph{a} & \emph{v} & \emph{av} & \emph{a} & \emph{v} & \emph{av} &   \emph{$a(A\neg V)$} &   \emph{$v(V\neg A)$} & \emph{a} & \emph{v} & \emph{av} & \emph{$\mu_A$} & \emph{$\mu_V$} & \emph{$\mu_{A \cap V}$} \\
\midrule
CAV-MAE & 20.24 & 24.82 & 29.98 & 40.72 & 39.20 & 47.43 & 25.09 & 23.72 & 67.85 & 56.40 & 67.33 & 4.20 & 4.78 & 0.81 \\
DeepAVFusion & 15.54 & 13.48 & 26.79 & 30.16 & 22.58 & 42.23 & 18.24 & 13.33 & 50.31 & 32.49 & 59.94 & 4.06 & 3.27 & 0.25 \\
Equi-AV & 17.44 & 13.83 & 25.03 & 34.41 & 22.91 & 39.78 & 21.03 & 13.27 & 57.34 & 32.96 & 56.47 & 7.41 & 5.97 & 1.42 \\
AV-Siam & 19.53 & 25.21 & 28.06 & 39.36 & 39.71 & 43.83 & 23.87 & 24.90 & 65.58 & 57.13 & 62.22 & 11.35 & 8.24 & 3.81 \\

\midrule

Gemini 1.5 Flash & 1.07 & 19.02 & 20.31 & 9.96 & 42.19 & 47.01 & 8.39 & 30.21 & 21.93 & 52.19 & 62.56 & 9.11 & 3.89 & 0.95 \\
Gemini 1.5 Pro & 2.02 & 28.10 & 29.48 & 14.46 & 55.57 & 58.50 & 10.50 & 32.85 & 24.83 & 74.26 & 78.56 & 2.39 & 4.30 & 0.42 \\
Gemini 2.0 Flash & 2.65 & 16.29 & 15.26 & 13.09 & 38.68 & 40.49 & 9.96 & 26.36 & 19.09 & 46.34 & 50.40 & 2.71 & 4.54 & 0.98 \\

\midrule

VideoLLaMA 2 & 17.94 & 24.00 & 27.77 & 41.58 & 50.04 & 53.53 & 27.03 & 37.94 & 58.35 & 52.35 & 58.94 & 15.64 & 5.18 & 3.05 \\
Unified-IO 2 & 15.89 & 15.17 & 30.00 & 36.48 & 31.18 & 50.34 & 24.01 & 22.13 & 50.91 & 32.42 & 62.88 & 8.85 & 4.74 & 1.79 \\
PandaGPT & 3.00 & 5.57 & 6.59 & 16.03 & 19.44 & 21.52 & 11.60 & 19.15 & 16.60 & 16.90 & 18.54 & 7.95 & 5.83 & 2.55 \\
OLA & 13.77 & 11.06 & 21.10 & 42.94 & 27.32 & 46.67 & 34.16 & 17.50 & 49.44 & 25.33 & 49.05 & 16.50 & 6.55 & 2.40 \\

\bottomrule
\end{tabular}
}
\caption{\textbf{Audio-visual video classification results on the subset of \vggsounder{} that is labelled as \textit{not} containing \textit{voice over} narrations}}\label{tab:no_voice_over}
\end{table*}

\noindent Considering the ``voice-over'' split (Tab.,\ref{tab:voice_over}) with its complement (Tab.,\ref{tab:no_voice_over}) exposes a two–way story that depends on how each model treats speech.

\vspace{0.7\baselineskip}
\noindent \textit{(i) Embedding models are confused.}  For all four embedding models, results when using \textit{audio} inputs jump by roughly $+5$–$10\%$ in $F_1$ when the narration track is removed, and Hit for audio climbs in parallel. This confirms that spoken commentary \emph{masks} class-specific sounds.
However, once silenced, the models can finally ``hear'' the underlying events, again, similar to the background music meta-class.

\vspace{0.7\baselineskip}
\noindent \textit{(ii) Reduced performance for speech-centric foundation models.}
Gemini 1.5 and PandaGPT fail when narration disappears: $F_1$ for audio inputs plunges by around \mbox{$ -17\%$} and Hit for audio inputs drops by up to 39\%. Our intuition is, that these models exploit the speech content as a shortcut. 

\vspace{0.7\baselineskip}
\noindent \textit{(iii) Middle ground for broad-coverage LMMs.}  Unified-IO 2 and VideoLLaMA-2 are between the two extremes: they register a moderate audio lift (+4–5\%) and a small visual bump (+1–2\%), yielding a $,+1$–$,8\%$ improvement in terms of $F_1$ score for audio-visual inputs. We hypothesise that their balanced training helps them survive the removal of speech while still profiting from the clearer acoustic scene.

\vspace{0.7\baselineskip}
\noindent \textit{(iv) Modality-confusion $\mu$ reacts in both directions.}  For speech-reliant models, clearer acoustics \emph{reduce} uncertainty, whereas for event-focused encoders (trained on \vggsound{}) it slightly \emph{rises} because the freshly revealing audio now dominates the fusion gate.

\vspace{0.7\baselineskip}
\noindent Taken together, voice-over narration acts as the mirror image of background music: it can be a \emph{helpful shortcut} for speech-aware foundation models, yet a \emph{destructive mask} for sound classifiers trained on \vggsound{}. 

\vspace{0.7\baselineskip}
\mypara{Confounder–free subset}

\begin{table*}[ht!]
\centering
\resizebox{\textwidth}{!}{
\begin{tabular}{l S[table-format=2.2] S[table-format=2.2] S[table-format=2.2] S[table-format=2.2] S[table-format=2.2] S[table-format=2.2] S[table-format=2.2] S[table-format=2.2] S[table-format=2.2] S[table-format=2.2] S[table-format=2.2] S[table-format=2.2] S[table-format=2.2] S[table-format=2.2]}
\toprule
& \multicolumn{3}{c}{\textbf{Subset Accuracy} \small$\uparrow$} & \multicolumn{5}{c}{$\mathbf{F_1}$ \small$\uparrow$} & \multicolumn{3}{c}{\textbf{Hit} \small$\uparrow$} & \multicolumn{3}{c}{$\mathbf{\mu}$ \small$\downarrow$} \\
\cmidrule(lr){2-4} \cmidrule(lr){5-9} \cmidrule(lr){10-12} \cmidrule(lr){13-15}
\textbf{Model} & \emph{a} & \emph{v} & \emph{av} & \emph{a} & \emph{v} & \emph{av} &   \emph{$a(A\neg V)$} &   \emph{$v(V\neg A)$} & \emph{a} & \emph{v} & \emph{av} & \emph{$\mu_A$} & \emph{$\mu_V$} & \emph{$\mu_{A \cap V}$} \\
\midrule
CAV-MAE & 20.33 & 24.37 & 29.63 & 41.53 & 39.69 & 47.94 & 21.52 & 22.14 & 69.17 & 58.20 & 69.31 & 3.82 & 5.10 & 0.92 \\
DeepAVFusion & 16.04 & 13.66 & 26.80 & 31.48 & 23.36 & 43.26 & 15.92 & 13.31 & 52.47 & 34.27 & 62.55 & 4.14 & 3.38 & 0.27 \\
Equi-AV & 17.64 & 13.86 & 25.05 & 35.47 & 23.53 & 40.73 & 18.23 & 13.04 & 59.08 & 34.51 & 58.88 & 7.27 & 6.33 & 1.57 \\
AV-Siam & 19.76 & 25.31 & 28.05 & 40.33 & 40.70 & 44.60 & 20.57 & 22.59 & 67.17 & 59.68 & 64.47 & 10.11 & 8.73 & 4.06 \\

\midrule

Gemini 1.5 Flash & 1.05 & 18.83 & 20.36 & 9.71 & 42.78 & 47.77 & 8.97 & 30.07 & 21.46 & 53.33 & 63.87 & 8.32 & 4.03 & 0.95 \\
Gemini 1.5 Pro & 1.98 & 27.41 & 29.10 & 14.28 & 55.72 & 58.80 & 10.79 & 31.65 & 24.25 & 74.88 & 78.97 & 1.75 & 4.44 & 0.31 \\
Gemini 2.0 Flash & 2.66 & 16.09 & 15.48 & 13.26 & 39.06 & 41.13 & 8.64 & 25.83 & 19.02 & 47.44 & 51.18 & 2.55 & 4.87 & 1.00 \\

\midrule
VideoLLaMA 2 & 18.00 & 23.76 & 27.29 & 41.95 & 50.45 & 53.79 & 23.90 & 37.03 & 59.33 & 53.69 & 60.41 & 13.78 & 5.49 & 3.14 \\
Unified-IO 2 & 16.10 & 14.63 & 29.79 & 37.96 & 30.59 & 50.62 & 22.10 & 21.37 & 53.55 & 32.83 & 64.96 & 8.96 & 4.90 & 1.94 \\
PandaGPT & 3.26 & 5.22 & 6.30 & 17.15 & 19.33 & 21.56 & 13.79 & 19.85 & 17.57 & 17.11 & 19.06 & 7.81 & 5.90 & 2.60 \\
OLA & 14.30 & 10.84 & 20.35 & 44.79 & 26.99 & 46.35 & 37.15 & 16.54 & 51.85 & 25.36 & 49.39 & 17.22 & 6.81 & 2.72 \\

\bottomrule
\end{tabular}
}
\caption{\textbf{Audio-visual video classification results on the subset of \vggsounder{} that is labelled as \textit{not} containing \textit{background music, static images, or voice over} narrations}}\label{tab:no_meta}
\end{table*}

\noindent The split that \emph{simultaneously} excludes background music, static images, and voice-over narration (Tab.,\ref{tab:no_meta}) serves as an upper-bound reference and reveals how each system performs when no major nuisance factor is present.

\vspace{0.7\baselineskip}
\noindent Removing \emph{all} three meta-classes unlocks the highest scores yet observed and sharpens modality agreement.

\subsection{Ablation study for additional labels in \vggsounder{}}
In this section, we conduct an ablation study to quantify the benefits introduced by different components of our annotation pipeline described in Section 3. Specifically, we compare model performance on three variants of ground-truth labels:
(a) Original \vggsound{} labels extended only with added synonymous and superclass labels, (b) Original \vggsound{} labels extended exclusively with human annotations, (c) Original \vggsound{} labels extended with human annotations and additional labels according to \cref{tab:class-mapping} (\vggsounder{}).

\noindent Detailed performance results in \cref{tab:vggsound_plus_heuristics} and \cref{tab:vggsound_plus_humans} demonstrate a consistent improvement across HIT and F1 metrics when employing the complete set of annotations (scenario c). This clearly illustrates the reduction in false-positive identifications and improved accuracy achieved through our annotation pipeline. This again highlights the importance of combining automated processes with thorough human verification in creating robust benchmarks for evaluating audio-visual models.

\begin{table*}[th!]
\centering
\resizebox{\textwidth}{!}{
\begin{tabular}{l S[table-format=2.2] S[table-format=2.2] S[table-format=2.2] S[table-format=2.2] S[table-format=2.2] S[table-format=2.2] S[table-format=2.2] S[table-format=2.2] S[table-format=2.2] S[table-format=2.2] S[table-format=2.2] S[table-format=2.2]}
\toprule
& \multicolumn{3}{c}{\textbf{Subset Accuracy} \small$\uparrow$} & \multicolumn{3}{c}{$\mathbf{F_1}$ \small$\uparrow$} & \multicolumn{3}{c}{\textbf{Hit} \small$\uparrow$} & \multicolumn{3}{c}{$\mathbf{\mu}$ \small$\downarrow$} \\
\cmidrule(lr){2-4} \cmidrule(lr){5-7} \cmidrule(lr){8-10} \cmidrule(lr){11-13}
\textbf{Model} & \emph{a} & \emph{v} & \emph{av} & \emph{a} & \emph{v} & \emph{av} & \emph{a} & \emph{v} & \emph{av} & \emph{$\mu_A$} & \emph{$\mu_V$} & \emph{$\mu_{A \cap V}$} \\
\midrule
Gemini 1.5 Flash & 0.31 & 22.12 & 23.60 & 1.71 & 33.15 & 35.94 & 2.98 & 31.83 & 41.23 & 1.51 & 4.17 & 0.09 \\
Gemini 1.5 Pro & 1.29 & 25.77 & 21.31 & 4.43 & 36.41 & 35.62 & 6.11 & 41.72 & 45.70 & 1.62 & 5.41 & 0.24 \\
Gemini 2.0 Flash & 5.70 & 20.29 & 19.39 & 9.95 & 32.34 & 33.91 & 9.49 & 30.55 & 35.31 & 2.50 & 4.77 & 0.63 \\

\midrule

VideoLLaMA 2 & 27.98 & 17.01 & 21.46 & 41.32 & 31.46 & 36.80 & 30.05 & 22.72 & 27.90 & 11.16 & 2.85 & 1.42 \\
Unified-IO 2 & 32.28 & 20.24 & 52.40 & 43.71 & 33.84 & 64.06 & 33.71 & 22.84 & 54.20 & 4.88 & 3.42 & 0.87 \\
PandaGPT & 5.20 & 7.65 & 8.95 & 12.68 & 16.83 & 19.55 & 8.54 & 11.23 & 13.30 & 4.51 & 4.48 & 0.94 \\
OLA & 10.71 & 8.63 & 14.29 & 23.33 & 17.81 & 28.86 & 18.06 & 10.95 & 22.41 & 7.61 & 4.05 & 0.71 \\

\bottomrule
\end{tabular}
}
\caption{\textbf{Audio-visual video classification results on \vggsound{}} inputs.}\label{tab:vggsound_perf}
\end{table*}

\begin{table*}[th!]
\centering
\resizebox{\textwidth}{!}{
\begin{tabular}{l S[table-format=2.2] S[table-format=2.2] S[table-format=2.2] S[table-format=2.2] S[table-format=2.2] S[table-format=2.2] S[table-format=2.2] S[table-format=2.2] S[table-format=2.2] S[table-format=2.2] S[table-format=2.2] S[table-format=2.2]}
\toprule
& \multicolumn{3}{c}{\textbf{Subset Accuracy} \small$\uparrow$} & \multicolumn{3}{c}{$\mathbf{F_1}$ \small$\uparrow$} & \multicolumn{3}{c}{\textbf{Hit} \small$\uparrow$} & \multicolumn{3}{c}{$\mathbf{\mu}$ \small$\downarrow$} \\
\cmidrule(lr){2-4} \cmidrule(lr){5-7} \cmidrule(lr){8-10} \cmidrule(lr){11-13}
\textbf{Model} & \emph{a} & \emph{v} & \emph{av} & \emph{a} & \emph{v} & \emph{av} & \emph{a} & \emph{v} & \emph{av} & \emph{$\mu_A$} & \emph{$\mu_V$} & \emph{$\mu_{A \cap V}$} \\
\midrule
Gemini 1.5 Flash & 2.97 & 25.19 & 26.28 & 11.86 & 42.60 & 46.37 & 21.94 & 45.57 & 56.77 & 11.62 & 4.16 & 0.87 \\
Gemini 1.5 Pro & 4.29 & 33.65 & 30.04 & 17.50 & 53.35 & 53.78 & 25.18 & 64.02 & 69.24 & 2.80 & 4.89 & 0.55 \\
Gemini 2.0 Flash & 4.16 & 22.29 & 20.52 & 12.70 & 40.06 & 41.13 & 14.58 & 41.95 & 46.01 & 2.48 & 4.94 & 0.82 \\

\midrule

VideoLLaMA 2 & 29.86 & 29.75 & 35.63 & 49.83 & 52.76 & 57.62 & 52.30 & 48.35 & 55.75 & 16.59 & 5.36 & 2.61 \\
Unified-IO 2 & 25.94 & 20.46 & 48.52 & 45.40 & 36.88 & 63.91 & 46.71 & 30.34 & 65.21 & 9.48 & 4.70 & 1.54 \\
PandaGPT & 5.13 & 6.49 & 7.65 & 19.28 & 19.43 & 21.72 & 14.89 & 13.52 & 15.42 & 7.67 & 5.31 & 2.26 \\
OLA & 19.93 & 13.57 & 22.53 & 48.02 & 29.27 & 46.50 & 45.70 & 22.38 & 43.74 & 15.14 & 6.36 & 2.15 \\

\bottomrule
\end{tabular}
}
\caption{\textbf{Audio-visual video classification results on \vggsound{} + human annotations}.}\label{tab:vggsound_plus_heuristics}
\end{table*}

\begin{table*}[th!]
\centering
\resizebox{\textwidth}{!}{
\begin{tabular}{l S[table-format=2.2] S[table-format=2.2] S[table-format=2.2] S[table-format=2.2] S[table-format=2.2] S[table-format=2.2] S[table-format=2.2] S[table-format=2.2] S[table-format=2.2] S[table-format=2.2] S[table-format=2.2] S[table-format=2.2]}
\toprule
& \multicolumn{3}{c}{\textbf{Subset Accuracy} \small$\uparrow$} & \multicolumn{3}{c}{$\mathbf{F_1}$ \small$\uparrow$} & \multicolumn{3}{c}{\textbf{Hit} \small$\uparrow$} & \multicolumn{3}{c}{$\mathbf{\mu}$ \small$\downarrow$} \\
\cmidrule(lr){2-4} \cmidrule(lr){5-7} \cmidrule(lr){8-10} \cmidrule(lr){11-13}
\textbf{Model} & \emph{a} & \emph{v} & \emph{av} & \emph{a} & \emph{v} & \emph{av} & \emph{a} & \emph{v} & \emph{av} & \emph{$\mu_A$} & \emph{$\mu_V$} & \emph{$\mu_{A \cap V}$} \\
\midrule
Gemini 1.5 Flash & 2.40 & 18.46 & 19.74 & 14.49 & 41.43 & 46.10 & 31.52 & 50.96 & 61.31 & 14.06 & 3.96 & 1.39 \\
Gemini 1.5 Pro & 3.60 & 27.33 & 26.95 & 19.69 & 54.75 & 57.06 & 34.51 & 72.99 & 78.22 & 3.25 & 3.96 & 0.61 \\
Gemini 2.0 Flash & 2.19 & 16.03 & 14.88 & 12.69 & 38.22 & 39.79 & 18.94 & 46.05 & 49.73 & 2.75 & 4.57 & 1.02 \\

\midrule

VideoLLaMA 2 & 15.46 & 23.78 & 27.57 & 39.91 & 49.91 & 53.34 & 56.86 & 52.24 & 58.61 & 15.75 & 5.30 & 3.05 \\
Unified-IO 2 & 14.05 & 15.01 & 29.35 & 35.71 & 31.03 & 49.72 & 51.36 & 32.02 & 61.62 & 10.56 & 4.65 & 1.75 \\
PandaGPT & 3.22 & 5.55 & 6.53 & 17.73 & 19.58 & 21.50 & 19.11 & 17.07 & 18.39 & 9.49 & 6.29 & 2.97 \\
OLA & 15.02 & 10.83 & 19.62 & 46.75 & 27.03 & 46.46 & 55.67 & 25.19 & 50.17 & 16.70 & 6.26 & 2.71 \\

\bottomrule
\end{tabular}
}
\caption{\textbf{Audio-visual video classification results on \vggsound{} + human annotations + automatically added labels}}\label{tab:vggsound_plus_humans}
\end{table*}

\subsection{\vggsounder{} Changelog}

\textbf{In October 2025}, we refined the definition of the modality confusion metric $\mu$ introduced in Section~\cref{sec:experiments}. In its current implementation, the $\mu$ metric marks a sample where a unimodal prediction is correct with respect to all unimodal labels, but the corresponding multimodal prediction is incorrect when evaluated against all available labels for that sample. Previously, multimodal predictions were compared only to the AV labels. For reference, we include the results obtained using the previous definition in the \cref{tab:main-old}. Old implementation code can be found via this link \url{https://pypi.org/project/vggsounder/0.1.4.1/}

\begin{table*}[!h]
\centering
\resizebox{\textwidth}{!}{
\begin{tabular}{l S[table-format=2.2] S[table-format=2.2] S[table-format=2.2] S[table-format=2.2] S[table-format=2.2] S[table-format=2.2] S[table-format=2.2] S[table-format=2.2] S[table-format=2.2] S[table-format=2.2] S[table-format=2.2] S[table-format=2.2] S[table-format=2.2] S[table-format=2.2]}
\toprule
& \multicolumn{3}{c}{\textbf{Subset Accuracy} \small$\uparrow$} & \multicolumn{5}{c}{$\mathbf{F_1}$ \small$\uparrow$} & \multicolumn{3}{c}{\textbf{Hit} \small$\uparrow$} & \multicolumn{3}{c}{$\mathbf{\mu}$ \small$\downarrow$} \\
\cmidrule(lr){2-4} \cmidrule(lr){5-9} \cmidrule(lr){10-12} \cmidrule(lr){13-15}
\textbf{Model} & \emph{a} & \emph{v} & \emph{av} & \emph{a} & \emph{v} & \emph{av} & \emph{a($A\neg V$)} & \emph{v($V\neg A$)} & \emph{a} & \emph{v} & \emph{av} & \emph{$\mu_A$} & \emph{$\mu_V$} & \emph{$\mu_{A\cap V}$} \\

\midrule[0.75pt]
\multicolumn{15}{l}{\emph{Embedding Models}}\\
\addlinespace[2pt] 

CAV-MAE & \bfseries 13.19 & 19.23 & \bfseries 24.49 & \bfseries 34.46 & 34.91 & \bfseries 42.62 & \bfseries 13.94 & \bfseries 19.00 & \bfseries 62.29 & 53.44 & \bfseries 64.17 & \bfseries 3.58 & 6.43 & 0.77 \\
DeepAVFusion & 10.19 & 11.10 & 21.53 & 25.31 & 21.29 & 37.35 & 10.37 & 10.55 & 45.77 & 32.61 & 56.27 & 3.74 & \bfseries 3.93 & \bfseries 0.17 \\
Equi-AV & 11.60 & 10.52 & 20.00 & 29.39 & 20.42 & 34.69 & 12.55 & 10.65 & 53.12 & 31.26 & 52.24 & 6.97 & 7.13 & 1.38 \\
AV-Siam & 12.79 & \bfseries 19.75 & 22.83 & 33.30 & \bfseries 35.41 & 39.43 & 12.90 & 18.21 & 60.19 & \bfseries 54.20 & 59.36 & 9.36 & 8.80 & 3.58 \\

\midrule
\multicolumn{15}{l}{\emph{Closed-source Foundation Models}}\\
\addlinespace[2pt]

Gemini 1.5 Flash & 1.78 & 14.44 & 16.44 & 14.49 & 36.98 & 42.52 & 15.61 & 21.61 & 32.73 & 47.36 & 59.10 & 10.22 & \bfseries 4.25 & 0.77 \\
Gemini 1.5 Pro & \bfseries 3.05 & \bfseries 20.86 & \bfseries 22.53 & \bfseries 19.26 & \bfseries 49.73 & \bfseries 53.74 & \bfseries 17.73 & \bfseries 22.90 & \bfseries 35.03 & \bfseries 69.23 & \bfseries 75.42 & \bfseries 2.09 & 4.85 & \bfseries 0.57 \\
Gemini 2.0 Flash & 1.85 & 12.54 & 12.69 & 11.80 & 34.08 & 36.45 & 6.19 & 18.90 & 18.51 & 43.83 & 47.72 & 2.39 & 5.43 & 1.00 \\

\midrule
\multicolumn{15}{l}{\emph{Open-source Foundation Models}}\\
\addlinespace[2pt] 

VideoLLaMA 2 & 12.86 & 19.85 & 24.47 & 38.87 & \bfseries 47.82 & \bfseries 52.35 & 20.34 & \bfseries 28.08 & \bfseries 58.91 & \bfseries 52.02 & 59.80 & 12.72 & 5.46 & 2.95 \\
Unified-IO 2 & 11.94 & \bfseries 11.56 & \bfseries 25.61 & 35.31 & 27.92 & 48.89 & 21.38 & 16.53 & 54.39 & 31.05 & \bfseries 65.11 & 8.70 & \bfseries 5.16 & \bfseries 1.79 \\
PandaGPT & 3.19 & 4.19 & 5.46 & 18.73 & 18.56 & 20.85 & 16.82 & 14.40 & 21.08 & 17.01 & 18.82 & \bfseries 7.59 & 5.90 & 2.47 \\
OLA & \bfseries 14.11 & 8.69 & 18.19 & \bfseries 47.70 & 24.85 &  46.48 & \bfseries 40.44 & 13.45 & 59.05 & 24.57 & 51.51 & 15.47 & 6.80 & 2.49 \\
\bottomrule
\end{tabular}
}
\caption{\textbf{Audio-visual video classification results on \vggsounder{}} for \vggsounder{} with the initial $\mu$ definition. }\label{tab:main-old}
\vspace{-0.7em}
\end{table*}
\FloatBarrier
\noindent \textbf{In June 2026}, we release a substantial label-quality pass over VGGSounder. Beyond re-annotating noisy samples, we also replace the original majority-vote label aggregation with a reliability-aware, seeded Dawid--Skene merge that models explicit negative votes, now part of the main pipeline discussed in \cref{sec:dawid-skene}.\\

\noindent In addition, an automated audit flagged 3,616 samples as potentially mislabeled. These were reviewed by human annotators (\cref{sec:human-lalbel-app}). Samples with no valid labels were removed. 

\noindent \\ All deletions, caused by manually annotated samples (3,616 samples) and label merging with Dawid-Skene, were automatically cross-checked using \texttt{gemini-3.5-flash} (\cref{sec:lvm-proposals-app}). \\

\noindent This resulted in $37395$ annotations in total (compared to $41173$ previously) with higher precision w.r.t. the gold standard. 

\end{document}